\documentclass[]{llncs}
\let\proof\relax
\let\endproof\relax
\usepackage{amsmath,amssymb,amsthm}
\usepackage{tikz}
\usepackage[utf8]{inputenc}
\usepackage[T1]{fontenc}
\usepackage{parskip} \usepackage{physics}
\usepackage{mathtools}
\usepackage{enumitem}
\usepackage[most]{tcolorbox}
\usepackage{hyperref}
\usepackage{qcircuit}
\usepackage{float}

\newif\ifshowdiscussion
\showdiscussiontrue \newcommand{\discussion}[1]{\ifshowdiscussion\textcolor{red}{~{#1}}\fi}
\newcommand{\discussionb}[1]{\ifshowdiscussion\textcolor{blue}{~{#1}}\fi}

\newtheorem*{landauer*}{Landauer's principle}
\newtheorem*{szil\'ard*}{Szil\'ard's engine}
\newtheorem*{rev*}{Reversible computation}
\newtheorem*{finetti*}{Finite de Finetti's Theorem}

\let\definition\relax
\spnewtheorem{definition}{Definition}{\bfseries}{}

\newcommand{\bigO}{\mathcal{O}}

\newcommand{\len}[1]{{\bf len}(#1)}
\newcommand{\negl}[1]{{\bf negl}(#1)}
\newcommand{\poly}[1]{{\bf poly}(#1)}
\newcommand{\spoly}[1]{{\bf superpoly}(#1)}
\newcommand{\nnegl}[1]{{ \textbf{non-negl}}(#1)}
\newcommand{\expp}[0]{{\bf exp}}
\newcommand{\ewp}[1]{\tag*{\scalebox{1}{\qquad\rm$[ \mbox{e.w.p.}\le #1]$}$\,.$}}
\newcommand{\ewpv}[1]{\tag*{\scalebox{1}{\qquad\rm$[ \mbox{e.w.p.}\le #1]$}$\,,$}}
\newcommand{\wnnp}[1]{\tag*{\scalebox{1}{\qquad\rm$[ \mbox{w.p.~} \nnegl{#1}]$}$\,.$}}
\newcommand{\wnnpv}[1]{\tag*{\scalebox{1}{\qquad\rm$[ \mbox{w.p.~} \nnegl{#1}]$}$\,,$}}
\DeclarePairedDelimiter{\ceil}{\lceil}{\rceil}
\DeclareMathOperator*{\argmin}{\arg\!\min}

\usepackage{epigraph}

\usepackage{etoolbox}

\makeatletter
\patchcmd{\epigraph}{\@epitext{#1}}{\itshape\@epitext{#1}}{}{}
\makeatother

\title{Key Agreement and Oblivious Transfer\\ from Free-Energy Limitations}
\author{Xavier Coiteux-Roy  \and Stefan Wolf}
\institute{Universit\`a della Svizzera italiana, Lugano, Switzerland.\\
\email{\{xavier.coiteux.roy,stefan.wolf\}@usi.ch}}
\date{\today}  

\begin{document}

\maketitle

\noindent
\makebox[\linewidth]{\small \today}

\begin{minipage}{1.01\linewidth}
~
\begin{abstract}
We propose one of the very few \emph{constructive} consequences of the second law of thermodynamics. More specifically, we present protocols for secret-key establishment and multiparty computation the security of which is based fundamentally on Landauer's principle. The latter states that the erasure cost of each bit of information is at least $k_{\rm B} \rm{T} \ln 2$ (where $k_{\rm B}$ is Boltzmann's constant and ${\rm T}$ is the absolute temperature of the environment). Albeit impractical, our protocols explore the limits of reversible computation, and the only assumption about the adversary is her inability to access a quantity of free energy that is exponential in the one of the honest participants. Our results generalize to the quantum realm.
\end{abstract}
\end{minipage}
\vspace{0.5cm}
\pagestyle{plain} 

{\bf Keywords:} Reversible computation, quantum information, information-\\theoretic security, key establishment, oblivious transfer.

\section{Introduction}
\subsection{Motivation}
In the past decades, several attempts were made to achieve cryptographic security from physical properties of communication channels: Most prominently, of course, \emph{quantum cryptography}~\cite{bb84,ekert1991quantum}; other systems made use of noise in communication channels~\cite{wyner1975wire} or bounds on the memory space accessible by an adversary~\cite{maurer1992conditionally}. These schemes have in common that no limit is assumed on the opponent's computational power: They are \emph{information-theoretically secure}. 

Our schemes for achieving confidentiality (key agreement or, more precisely, \emph{key expansion}) as well as secure co\"operation (multiparty computation, \emph{i.e.}, \emph{oblivious transfer}) rely solely on a bound on the accessible \emph{free energy}\footnote{Free energy is ``free'' in the sense that it can be used to do work --- it is not ``entrapped'' in a system.} of an adversary. More specifically, we propose schemes the security of which follows from \emph{Landauer's principle}, which is a quantification of \emph{the second law of thermodynamics}: \emph{In a closed system, ``entropy'' does not decrease} (roughly speaking). 

\emph{Landauer's principle} states that the \emph{erasure of information} unavoidably costs free energy, the amount of which is proportional to the length of the string to be erased. On the ``positive'' side, the \emph{converse} of the principle states that the all-$0$ string of length $N$ has a free-energy value proportional to $N$. More precisely, the erasure cost and work value are both quantified by $k_{\rm B} {\rm T}\ln 2\cdot N$, where $k_{\rm B}$ is \emph{Boltzmann's constant} (in some sense the nexus between the micro- and macroscopic realms), and ${\rm T}$ is the absolute temperature of the environmental heat bath. 

Our result can be seen as one episode in a series of results suggesting information-theoretic security to be, in principle, achievable under the assumption that \emph{at least one in a list of physical theories}, such as quantum mechanics or special relativity, \emph{is accurate}: We add to this list the second law of thermodynamics~--- to which not much glamour has been attached before. 

\subsection{Contributions}
We base the ``free-energy-bounded model'' of information-theoretic cryptography upon the observation that the second law of thermodynamics has a cryptographically useful corollary: ``Copying information has a fundamental cost in free energy.'' Bounding the free energy of an adversary forces them into picking parsimoniously what to copy, and that can be exploited in a reversible-computing context to ensure information-theoretic security. Our secret-key establishment protocol demonstrates how bounds in free energy can lead to cryptographic mechanisms similar to the ones used in quantum-key distribution and in the bounded-storage model, while our oblivious-transfer protocol exemplifies the novelty of our model.

This is an overview of our article: In Section~\ref{stateoftheart}, we review the subjects of information-theoretic cryptography and of reversible computing. In Section~\ref{sectionmodel}, we introduce, based on reversible computing, a novel model of computation and interaction that captures the consumption and the production of free energy in Turing machines. In Section~\ref{secproof}, we establish some prerequisites: we prove a version of Landauer's principle in our framework, and construct a game that is basically equivalent to a thermodynamical ``almost-no-cloning theorem,'' which we later use in our security proofs. In Sections~\ref{secske}~and~\ref{secot}, we offer protocols for \emph{secret-key establishment} and \emph{oblivious transfer}, respectively; their information-theoretical security is based fundamentally on Landauer's principle. It is assured against adversaries whose bound in free energy is exponential compared to the one of the honest players. While the present work focuses on classical information, we sketch in Section~\ref{toquantum} how all our results generalize in presence of quantum adversaries.

\section{State of the Art}\label{stateoftheart}
\subsection{Information-theoretic cryptography from physical assumptions}

In parallel to the development of computationally secure
cryptography~--- and somewhat in its shadow~---, attempts were made to obtain in a provable fashion stronger, \emph{information-theoretic
   security},
based not on the hardness of obtaining the (uniquely
determined) message in question, but on the sheer lack of information. 
Hereby, the need for somehow 
``circumventing'' Shannon's pessimistic theorem of perfect secrecy is met by some sort of
\emph{physical 
limitation}. The latter can come in the form of simple noise in a
communication channel, a limitation on accessible memory, the
uncertainty principle of quantum theory, or
the 
non-signalling postulate of special relativity.

The first system of the kind, radically improving on the perfectly
secret
yet impractical \emph{one-time pad}, has been 
\emph{Aaron Wyner}'s wiretap channel~\cite{wyner1975wire}: 
Here, information-theoretic secret-key establishment becomes possible ---
under the assumption,
however, that the legitimate parties already start with an advantage, 
more specifically, that the adversary only has access to a
non-trivially
degraded version of the recipient's pieces of information. 
A \emph{broadcast scenario} was proposed by \emph{Csisz\'ar} and
\emph{K\"orner}~\cite{csiszar1978broadcast}~--- where, again, an initial advantage 
in terms of information proximity or information quality was required by the 
legitimate partners \emph{versus} the opponent. A~breakthrough 
was marked by the work of \emph{Maurer}~\cite{maurer1993secret}, who showed that the need
for such an initial advantage on the information level can be replaced
by \emph{interactivity} of communication: Maurer, in addition, 
conceptually simplified and generalized the model by separating 
the noisily correlated data generation from public yet
authenticated
communication, the latter being considered to be for free. 
The model shares its communication setting with both  
\emph{public-key} as well as \emph{quantum cryptography}. 
Maurer and Wolf~\cite{maurer1996towards} have shown that in the 
case of independent-channel access to a binary source, key agreement
is in fact possible in principle in \emph{all} non-trivial cases,
\emph{i.e.}, even when Eve starts with a massive initial advantage 
in information quality. 

In the same model, it has also been shown that \emph{multiparty
  computation} becomes possible, namely \emph{bit commitment}
and (the universal primitive of) \emph{oblivious transfer}~\cite{crepeau1988achieving,crepeau1997efficient}.
More generally, oblivious transfer has also been achieved from {\em
  unfair} 
noisy channels, where the error behaviour is prone to be influenced in
one way or another by the involved, distrusting parties willing to co\"operate. 

The \emph{public-randomizer model} by Maurer~\cite{maurer1992conditionally}
has generally been recognized as the birth of the idea of
``memory-bounded models,'' based on the fact that the \emph{memory} an 
opponent or cheater (depending on the context) can access is
limited. Specifically, Maurer assumes the wire-tapper can obtain a
certain \emph{fraction} of the physical bits. This was generalized to arbitrary \emph{types} of information
by \emph{Dziembowski} and Maurer~\cite{dziembowski2002tight}.
Analogously, also \emph{oblivious transfer} has been shown achievable 
with a memory-bounded receiver~\cite{cachin1998oblivious,ding2004constant}.
The main limitation to the memory-bounded model, for both secret-key establishment and multiparty computation, is that the memory advantage of the honest participants over the adversaries is at most quadratic~\cite{dziembowski2004generating}.

The idea to use \emph{quantum physics} for cryptographic ends dates
back 
to \emph{Wiesner}, who, for instance, proposed to use the uncertainty
principle 
to realize unforgeable banknotes. His ``conjugate coding''~\cite{wiesner1983conjugate} resembles
oblivious 
transfer; the latter~--- even bit commitment, actually~--- we know now to be unachievable from quantum
physics only~\cite{mayers1997unconditionally,lo1998quantum}. A breakthrough has been the now famous ``BB84''
protocol 
for key agreement by communication through a channel allowing for 
transmitting quantum bits, such as an optic fibre, plus a public yet
authenticated classical channel~\cite{bb84}. 

A combination of the ideas described is the ``bounded quantum-storage
model'' \cite{damgaard2008cryptography}: 
Whereas no quantum memory is needed at all for the honest players, a
successful adversary can be shown to need more than $n/2$ of the
communicated quantum bits.
The framework has been unified and generalized to the ``noisy'' model
by \emph{K\"onig}, \emph{Wehner}, and \emph{Wullschleger}~\cite{konig2012unconditional}.

Very influential has been a proof-of-principle result
by \emph{Barrett, Hardy, and {\linebreak}Kent}~\cite{barrett2005no}: The
security in key agreement that stems from witnessing quantum correlations can be established regardless of the validity of quantum theory, only from the postulate of special relativity that there is \emph{no superluminal signalling}. The authors combined \emph{Ekert}'s~\cite{ekert1991quantum}
idea to obtain secrecy from proximity to a pure state, guaranteed by 
\emph{close-to-maximal violation of a ``Bell inequality,''} with the 
role this same ``nonlocality'' plays in the argument that the
outcomes
of quantum measurements are, in fact, random and not predetermined: 
In the end, reasoning results that are totally \emph{independent} of the completeness of quantum 
theory. Later, efficient realizations of the paradigm were
presented~\cite{hanggi2010efficient,masanes2011secure}. 
Conceptually, an interesting resulting statement is that
information-theoretic 
key agreement is possible if \emph{either quantum mechanics OR
  relativity theory} are complete and accurate
``descriptions of nature.''  
Another point of interest is that trust in the manufacturer is not
even required: ``device independence''~\cite{PhysRevLett.113.140501}.

\emph{Kent} also demonstrated that bit commitment can be information-theoretically secure thanks to special relativity alone~\cite{kent1999unconditionally}. On the other hand, oblivious transfer cannot be information-theoretically secure even when combining (without further assumptions) the laws of quantum mechanics and special relativity~\cite{colbeck2007impossibility}.

\subsubsection{Now --- the free-energy-bounded model:}
We add to this list the novel \emph{free-energy-bounded model}.
Unlike the assumptions in memory-bounded models, thermodynamics does not in principle prohibit free-energy-bounded players from computing on memories of exponential size (in some security parameter), but it \emph{does} prohibit those players from \emph{erasing} a significant portion of such memories. If the players only have access to memories in \emph{initial states of maximal entropy}, as is assumed in equilibrium in thermodynamics, the erasing restriction becomes a \emph{copying} restriction (because one cannot copy without a blank memory to write the copy onto) and opens the way to a novel foundation of physics-based information-theoretic security that is different from the bounded-storage model.\footnote{In particular, the free-energy-bounded model offers fresh mechanisms, coming from reversible computing, to build information-theoretic protocols (e.g., our oblivious-transfer protocol). Another important difference is that in our protocols, the advantage of honesty in free-energy consumption is exponential in the security parameter, while in the bounded-storage model (which is not based on reversible computing but arguably more practical), it is polynomial.}

\vspace{-0.15cm}
\subsection{Reversible computing}\label{sectionthermo}
\vspace{-0.05cm}
\subsubsection{The cost of computation.}
Security in cryptography relies on a cost discrepancy between honest and malicious actors. While fundamental thermodynamical limits to the cost of computation have been well-studied (for example, see~\cite{faist2015minimal} for a quantum-informational analysis and~\cite{baumeler2019free} for an algorithmic-information-theoretical analysis), they have never before\footnote{Let us mention the (questionable) conjecture in \cite{hungerbuhler2003one} that the heat-flow equation of thermodynamics is a computational one-way function.} been considered as a means for cryptography --- we address that.

\vspace{-0.05cm}
\subsubsection{The second law of thermodynamics.}

The modern view of the second law of thermodynamics is due to \emph{Ludwig Boltzmann}, who defined \emph{the entropy of a macrostate}~--- roughly speaking, the natural logarithm of the number of microstates in the
macrostate in question~--- and stated that the entropy of a closed system does not decrease with time. The second law has constantly been subject to discourse, confusion, and dispute; its most serious challenge was ``\emph{Maxwell's demon}'' who apparently violates the law by adaptive acts, \emph{i.e.}, by a sorting procedure.
\emph{Charles Bennett}~\cite{bennett1987demons} explained that Maxwell's paradox actually disappears when the demon's internal state (its ``brain'') is taken into consideration. More specifically, \emph{the erasure} of the stored information requires free energy that is then dissipated as heat to the environment. This is \emph{Landauer's principle}~\cite{landauer61}; it did not only help to resolve the confusion around Maxwell's demon, but turned out to be an important manifestation of the second law with respect to information processing in its own right: Erasure of information~--- or, more generally, any logically irreversible computing step, has a  thermodynamic cost. \emph{Logical} irreversibility (information is lost)
implies \emph{thermodynamic} irreversibility (free energy is ``burnt'' to heat up the environment).\\

\begin{tcolorbox}[title=Landauer's principle.]
Erasing $n$ random bits requires to transform at least $n \cdot k_{\rm B} {\rm T} \ln 2 ~{\rm J}/{\rm K}$ of free energy into heat, which is dissipated into the environment.
\end{tcolorbox}

\subsubsection{Energy-neutral (thermodynamically reversible) computation.}
Landauer's principle serves as a strong motivation to ask for the possibility whether computing
can always be (made) \emph{reversible}, \emph{i.e.}, forced to not ``forget'' along the way any information about
the past (previous computation). More specifically, can every Turing-computable function
also be computed by a reversible Turing machine~(the latter was introduced in~\cite{lecerf1963machines}; see Chapter~5~of~\cite{morita2017theory} for a more modern account)? In the early 1970s, \emph{Charles Bennett}
answered this question to the affirmative; the running time is also at most
doubled, essentially~--- a very encouraging result~\cite{bennett73}: The imperative
reversibility of microphysics can, at least in principle, be carried over to macrocomputing. Bennett's
idea was that the reversible Turing machine would allocate part of its tape to maintain a history of its computation. While the latter
needs to be gotten rid of in order to have the whole be ``sustainable,'' that cannot be done by
``crude'' erasure of that history~--- all won would be lost again. It can, however, be done by \emph{un-computing}: After
copying the output, the reversible Turing machine reverts step by step the original computation, undoing its history tape in a
``controlled'' and reversible way until the output is computed back to the input. An idea similar to
Bennett's elegant trick also works for circuits: Any irreversible circuit can be transformed into
a reversible one, computing the same function, and having essentially only double depth.

All in all, this means that logical reversibility~--- which Landauer tells us to be a \emph{necessary}
condition for thermodynamic reversibility~--- can be achieved; remains the question whether it is also
a \emph{sufficient} condition for energy-neutral computation. The answer is \emph{yes}, as exemplified
by \emph{Fredkin and Toffoli}~\cite{fredkin1982conservative} and their \emph{Gedankenexperiment} of a ``ballistic computer''
which carries out its computations through elastic collisions between balls and balls, and balls and walls.

In the end, we get an optimistic picture for the future of computing: \emph{Any computable function can be computed also without
the transformation of free energy into heating of the environment}. (Clearly, a ``loan'' of free energy is necessary to start the computation, but no law of physics prevents its complete retrieval, alongside the result of the computation, when the latter concludes.)\\

\begin{tcolorbox}[title=Reversible computing.]
Any logically reversible computation can be done at zero free-energy cost by a reversible Turing machine.
\end{tcolorbox}

Reversible computing is at the core of our model.\footnote{Reversible computing is of paramount importance in the context of Moore's and Koomey's laws about the future of computation, because their continuation is threatened by physical walls and the most important one comes from thermodynamics (and not quantum mechanics). Reversible computing can in principle solve the problem completely by enabling computation without dissipation of heat. }

\subsubsection{The energy value of redundancy.}\label{szilardsection}
The converse of Landauer's principle states that all physical representations of the all-0 string have work value. More generally, all redundant (i.e., compressible in a lossless fashion) strings have work value, which is essentially their length minus their best compression~\cite{bennett1982thermodynamics}. A bound in free energy is therefore a bound on the redundancy of information; a principle we use in this work to construct cryptographic protocols.

\begin{figure}[htbp]
\begin{center}
\includegraphics[width=1\textwidth]{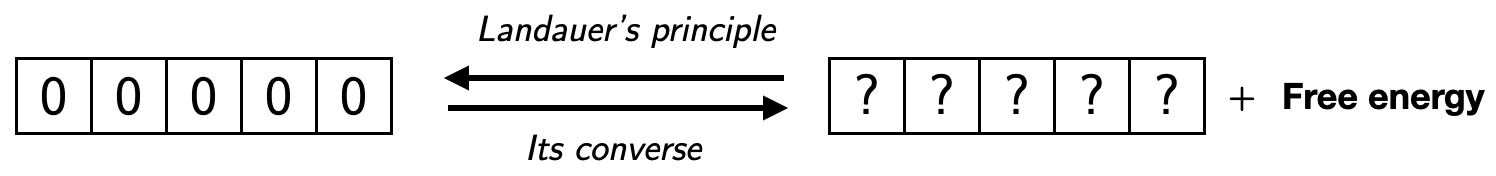}
\caption{Given the existence of thermodynamical heath baths, there is a fundamental equivalence between free energy and redundancy (\emph{i.e.}, the absence of randomness).} \label{figure1}
\end{center}
\end{figure}

\begin{tcolorbox}[title=The converse of Landauer's principle.]
It is possible to extract an amount $n \cdot k_{\rm B} T \ln 2$ of free energy from an environment by randomizing $n$ blank bits.\end{tcolorbox}

In the light of Landauer's principle and of its converse, the all-0 string can be used as a proxy for free-energy (see Fig.~\ref{figure1}). This allows us to abstract the thermodynamics completely from the model we present in Section~\ref{sectionmodel}, which is then formulated purely in terms of (logically reversible) Turing machines.

\section{Turing Machines with Polynomial Free-Energy Constraints}\label{sectionmodel}
In the following, we have this classical\footnote{The classical setting is used for all sections but Section~\ref{toquantum}, which approaches the quantum generalization.} setting in mind: Alice, Bob, and Eve have their own secure labs, where they can store and manipulate exponentially long (in some security parameter $\nu$) bit strings. Those strings start in uniformly random\footnote{This randomness is motivated by the equipartition assumption of classical thermodynamics.} states; we can think of them as the information about the specific microstate that describes the position and momentum of an exponential number of particles floating in their labs. We assume that technology is advanced enough to consider these exponentially long bit strings as static (even if the system starts in a random state, it does not get re-randomized at every time step), either because their evolution is tractable (it evolves according to the logically reversible laws of physics) or because the players can act on them quickly enough that it does not matter. The physical restriction on the honest and malicious players concerns their available free energy: For some security parameter $\nu$, malicious players are bounded exponentially (more precisely, by $2^\nu$), while honest players need only an asymptotically $\bigO(\nu)$ amount. These bounds are constraining because any computation that is not logically reversible has a free-energy cost; a malicious agent cannot for example erase a $2\cdot 2^{\nu}$-long segment of random information --- by Landauer's principle, doing so would cost a quantity of free energy exceeding their free-energy bound. We formalize this computation model in Section~\ref{computationmodel}.

Communicationwise, the players are allowed to broadcast $\bigO(\nu)$-length bit strings in the traditional sense using a public authenticated channel, or to transfer $\bigO(2^\nu)$-long bit strings through a private-but-insecure\footnote{By ``insecure,'' we mean here that it is vulnerable to Eve-in-the-middle attacks.} $\textsc{SWAP}$ channel, This channel, which swaps two bit strings at no energy cost, can also be substituted by an insecure \emph{physical} channel. Both views are informationally equivalent, and are defined in Section~\ref{reversiblecomm}.

In particular, our model differs from the bounded-storage
model --- both the players and the adversary have more power.

\subsection{Computation model}\label{computationmodel}
The fundamental laws of physics are logically reversible. We hence base
our formal notion of player (or adversary) on reversible Turing machines.\newcommand{\pfe}[0]{{\rm\textbf{PFE}}}
\newcommand{\bfe}[0]{{\rm\textbf{BFE}}}

\begin{definition}[TTM]\label{TTM}
A \emph{thermodynamical Turing machine} (TTM) is a logically reversible, deterministic, universal, prefix-free Turing machine with the following semi-infinite tapes:
\begin{enumerate}\rm
\item An input-only \textbf{instruction} tape.
\item An initially blank \textbf{computation} tape that must be returned blank when the machine halts.
\item An initially random \textbf{memory} tape.
\item An initially blank \textbf{free-energy} tape.
\end{enumerate}
\end{definition}

The \textbf{free-energy} tape of a TTM imitates a ``reservoir'' of free energy:

\begin{definition}[consumption]
 The \emph{free-energy input} $w_{\rm in}$ is quantified\footnote{More precisely, it is bounded from below.}, when the machine halts, by the distance, on the initially blank \textbf{free-energy} tape, between the extremity and the last cell with a $1$ (after this cell, the tape contains only 0s). 
\end{definition}

For example, if a machine always manages to return the \textbf{free-energy} tape as blank as it was --- it uses no free energy and computes both logically and thermodynamically reversibly; if a machine writes, and leaves, some information on the first $n$ cells of the initially blank \textbf{free-energy} tape, we say it \emph{consumes} an amount $w_{\rm in}=n$ of free-energy. ({In this work we have set $k_{\rm B} {\rm T} \ln2 \coloneqq 1$.})

Our security proofs will rely on a concept we name \textbf{proof-of-work}.
\begin{definition}[production]
We say a TTM produces a \emph{\textbf{proof-of-work}} of value $w_{\rm out}$ if it halts with a number $w_{\rm out}$ of $0$s at the beginning of its (initially random) \textbf{memory} tape.
\end{definition}

We consider agents (TTMs) with bounds, in the security parameter $\nu$, on the free-energy input.

\begin{definition}[\bfe]\label{def:pfe}
An $f(\nu)$-\bfe~agent --- an agent who is \emph{bounded in free energy by the function $f(\nu)$}, where $\nu$ is a security parameter --- is modelled by a TTM that can only consume a quantity $f(\nu)$ of free energy.
\end{definition}
In other words, every time a $f(\nu)$-\bfe~agent reaches a halting state, the non-blank portion of its \textbf{free-energy} tape ends at a distance at most $f(\nu)$ from the extremity, by definition.

In our protocols, the honest players are asymptotically $\bigO(\nu)$-\bfe, while the adversary is assumed exactly $2^{\nu}$-\bfe.
An important limitation of $f(\nu)$-\bfe~agents is given by the following theorem, to which the security of our protocols will be reduced.
\begin{theorem}\label{thm:violation}
For all $k>0$, an $f(\nu)$-\bfe~player cannot produce an $f(\nu)+k$ \textbf{proof-of-work}, except with probability $2^{-k}$.
\end{theorem}

The theorem is a consequence of the logical-reversibility characteristic imposed by the second law of thermodynamics. The proof is done in Section~\ref{app:violation}, based on Definitions~\ref{TTM}~and~\ref{def:pfe} (\emph{i.e.}, with no further references to thermodynamics).

\subsection{Communication and reversible transfer}\label{reversiblecomm}
Our cryptographic model can be formalized further by integrating \bfe~parties into a multi-round interactive protocol that uses reversible computing, however, let us focus on how Alice and Bob can exchange information. There are of two distinct resources: 
\begin{itemize}\rm
\item Standard communication for messages of length $\bigO(\nu)$.
\item Reversible transfer for longer messages, up to length $\bigO(2^\nu)$.
\end{itemize}

\subsubsection{Standard communication.}
We consider that Alice and Bob have access to a \emph{public} \emph{authenticated} communication channel in the traditional sense: Alice broadcasts a message (making, therefore, inevitably many copies of its information content) and Bob receives it. Because Alice and Bob are $\bigO(\nu )$-$\bfe$, this information-duplicating channel can only be used for messages of length $\bigO(\nu)$.

\subsubsection{Reversible transfer.}
To send states of length more than $\bigO(\nu )$, Alice and Bob have to resort to reversible computing. Reversible transfer differs from standard communication in the sense that, in order to implement the process at no free energy cost, the sender \emph{must forget} the information content of the message they send. (They could, of course, pre\"emptively make a partial copy of that information, but copying is not free and is thus limited by the free energy assumption.)
There are two different physical ways to picture such reversible transfer.

The first way is to implement, over a given distance, a reversible \textsc{SWAP}: In essence, this operation simply swaps two bit strings of equal length in a logically and thermodynamically reversible way --- Alice gets Bob's string and Bob gets Alice's string. Since we are only interested in the string that Alice (the sender) sends, Bob (the receiver) can input junk in exchange. The \textsc{SWAP} allows $\bigO(\nu )$-$\bfe$ players to transfer between themselves $\bigO(2^\nu)$ bits of information (without copying them).

The second way to implement reversible communication is to simply consider that Alice is sending the whole physical system encoding her string (\emph{e.g.}, she puts a canister of gas with entropy $2^\nu$ on a frictionless cart and pushes it toward Bob). For the cart as for the \textsc{SWAP} channel, since the information is never copied, it can be transferred from Alice to Bob at no thermodynamical cost. This is not dissimilar to how it is in practice cheaper to send hard drives directly by mail rather than to send their content through a cable.

These two pictures (the \textsc{SWAP} channel and the physical channel) are from an information point of view equivalent --- we adopt the \textsc{SWAP} channel for this work.

\section{Technical Preliminaries}\label{secproof}
We introduce some notation and introduce some of the techniques used later in the security proof of our main protocols.

\subsection{Smooth min-entropy}

Most of our formal propositions rely on the \emph{variational distance}. 

\begin{definition}
The \emph{variational distance} between two random variables $X$ and $Y$ is defined as 
\begin{equation}
\delta(X,Y)\coloneqq \frac{1}{2} \sum_{i \in \mathcal{X}\cup \mathcal{Y}} \left|p(X=i)-p(Y=i)\right|\,.
\end{equation}
It is operationally very useful because it characterizes the impossibility to distinguish between $X$ and $Y$ --- using any physical experiment whatsoever. More precisely, given either $X$ or $Y$ with probability $1/2$, the optimal probability to correctly guess which one it is is $(1+\delta(X,Y))/2$.
\end{definition}

\begin{definition}
 The \emph{conditional min-entropy} $H_\infty(X|Y)$ is defined as
\begin{equation} H_\infty (X|Y)\coloneqq -\log \sum_y P(Y=y)  \max_x  P(X = x|Y=y)\,.\end{equation}
\end{definition}
It is the optimal probability of correctly guessing $X$ given side information $Y$.

Smoothing entropies~\cite{renner2004smooth,renner2005simple} is done to ignore events that are typically unlikely. We will typically use smoothing with a parameter $\epsilon= \negl{\nu}$. We denote by $\negl{\nu}$ the functions that are \emph{negligible} in $\nu$, meaning asymptotically bounded from above by the inverse of every function that is polynomial in $\nu$.
\begin{definition}
The \emph{smooth conditional min-entropy} $H^\epsilon_\infty (X|Y)$ is defined as

\begin{equation}
H^\epsilon_\infty (X|Y) \coloneqq \max_{\omega \in \Omega \mbox{~s.t.~}P(\omega)\ge1-\epsilon} \min_y \min_x ( - \log P(X=x|Y=y,\omega)  ) \,,
\end{equation}
where $\Omega$ is the set of all events.
\end{definition}
Smooth conditional min-entropy is used mainly for privacy amplification.

\subsection{Proof of Theorem~\ref{thm:violation}}\label{app:violation}
We define and prove formally a version of Landauer's principle (Theorem~\ref{thm:violation}), which is the claim in Section~\ref{sectionmodel} that $\bfe$ players modelled as thermodynamical Turing machines cannot produce more free energy than they consume, except with exponentially vanishing probability. The theorem follows from the logical reversibility of a TTM --- the existence of a thermodynamically free logically irreversible physical process would be a violation of the second law of thermodynamics. We introduce some algorithmic-information-theory notation along the way; a more exhaustive introduction is the excellent book by Li and Vit\'anyi~\cite{li2008introduction}.

\setcounter{theorem}{0} \begin{theorem}[technical]\label{secondlaw}
Given infinite tapes $\{x,y\}$, a $f(\nu)$-\bfe~TTM $U_p(x,y)$ cannot produce a $f(\nu)+k$ \textbf{proof-of-work}, except with probability $2^{-k}$.
\end{theorem}

$\{p,x,y\}$ are, respectively, the representation of the \textbf{instruction}, \textbf{memory}, and (blank) \textbf{free-energy} tapes, at the beginning of the computation.

We start with the simpler case of assuming that all of these tapes are finite (but arbitrarily long), and then generalize our analysis to the infinite case.

\subsubsection{The finite case.}

Let $U_p(x,y)$ be a thermodynamical Turing machine as described in Definition~\ref{TTM}:  universal, prefix-free, deterministic and logically reversible. The program $p$ is taken from the read-only \textbf{instruction} tape (which can be taken long but finite); the (initially random) \textbf{memory} tape starts in $x\in_R\{0,1\}^{\len{x}}$, with $\len{x}$ taken arbitrary but finite; the \textbf{free-energy tape} starts with blank content $y=0^{\len{y}}$, where $\len{y}$ is also finite.

The logical-reversibility condition means $U_p(x,y)=U_p(x',y')$ if and only if $(x,y)=(x',y')$.

We use a counting argument. We consider the set $S$ of all couples $(x,y)$ of lengths fixed. There are $\#S=2^{\len{x}}$ of them and they are all equally probable. We then consider the subset
\begin{equation}
S(w_{\rm in},w_{\rm out}) \coloneqq \left\{ x,y {\rm ~s.t.~} U_p(x,y)=\tilde{x},\tilde{y}{\rm ~with~} \begin{cases}
\tilde{x}=0^{w_{\rm out}}\,||\,*\\
\tilde{y}=*\,||\,0^{\len{y}-w_{\rm in}}
\end{cases}\right\}\,,
\end{equation}
where $*$ is an arbitrary padding string of appropriate length, and $\,||\,$ denotes a concatenation.
Intuitively, $w_{\rm in}$ bounds the free-energy input and is the minimum number of bits that get randomized on the initially blank \textbf{free-energy} tape $y$; $w_{\rm out}$ bounds the free-energy output and is the maximum number of erased bits on the initially random \textbf{memory} tape $x$. (Those erased bits constitute the \textbf{proof-of-work}.)

\begin{lemma}
\begin{equation}
\# S(w_{\rm in},w_{\rm out})\le 2^{\len{x}-{w_{\rm out}}+{w_{\rm in}}}\,.\end{equation}
\begin{proof}
Because of logical reversibility, the input-couples $(x,y)\in S$ are at most\footnote{``At most'' because not all programs halt and some output-couples might not be in the image of $U_p$.} as numerous as the output-couples 
$
(\tilde{x},\tilde{y}) \mbox{~s.t.~} \begin{cases}
\tilde{x}=0^{w_{\rm out}}\,||\,*\\
\tilde{y}=*\,||\,0^{\len{y}-w_{\rm in}}
\end{cases}
$
\hspace{-12pt}. We count the maximum number of such output-couples by summing the lengths of all ``$*$ positions''; there are at most $2^{(\len{x}-w_{\rm out})+w_{\rm in}}$ of them.\qedhere
\end{proof}
\end{lemma}

The probability of drawing at random such a couple $(x,y)$ is therefore
\begin{equation}P(x,y\in S(w_{\rm in},w_{\rm out}))\le \# S(w_{\rm in},w_{\rm out}) / \#S=2^{w_{\rm in}-w_{\rm out}}\,.\end{equation}

\begin{proposition}\label{proposition6}
Given finite $\len{x}$ and $\len{y}$, a $f(\nu)$-\bfe~TTM $U_p(x,y)$ (therefore with free-energy input $w_{\rm in}=f(\nu)$) is limited in its production of free energy $w_{\rm out}$ by
\begin{equation}
\forall k>0,~ P\left( w_{\rm out} > w_{\rm in}+k \right) \le 2^{-k} \,.\end{equation}
\end{proposition}

\subsubsection{The infinite case.}
We now reduce the infinite case to the finite case that we just analyzed.

We take again a TTM. Let us consider $x\in_R\{0,1\}^\infty$, where each bit is perfectly random. Let us also set $y=0^\infty$. Since $p$ is fixed, it is enough to again consider it finite. The prefix-free condition implies that the behaviour of $U_p(x,y)$ is well defined even on infinite tapes because its programs\footnote{``Program'' is taken here in the general sense and includes arguments $p$ and $x$.} are self-delimited.

\newcommand{\prefix}[1]{\textit{\rm effective}(#1)}

\begin{definition}
Let \begin{equation}\displaystyle \Omega_{U_p}\coloneqq \sum_{\prefix{x} \text{~s.t.~} U_p(x,y) \text{~halts}} 2^{- \len{\prefix{x}}} \end{equation} be the \emph{halting probability} of $U_p$ (\emph{i.e}, Chaitin's constant~\cite{chaitin1975theory}), where the sum is over all self-delimited programs $\prefix{x}\in \{0,1\}^*$ \footnote{We assume $p$ to be fixed; by ``program'' we mean the random input $x$.}.
\end{definition}

We also define its partial sum.
\begin{definition}
\begin{equation}
\displaystyle{\Omega_{U_p}}(n)\coloneqq \sum_{\prefix{x} \text{~s.t.~} U_p(x,y) \text{~halts~and~} \len{\prefix{x}}\le n } 2^{- \len{\prefix{x}}}.\end{equation} 
\end{definition}

Note first that since ${\Omega_{U_p}}(n)$ is a monotonically increasing function that converges to ${\Omega_{U_p}}$, it holds that
\begin{equation}
\forall \epsilon > 0, \exists N' {\rm ~s.t.~} \Omega_{U_p}- \Omega_{U_p}(N') < \epsilon\,. \label{mono}\end{equation}

\begin{definition}
Let ${\rm BB}_{U_p}(n)$ be the \emph{time-busy-beaver function}, which returns the maximum running time that a halting program $\prefix{x}$ of length $\le n$ can take before halting.\end{definition}
Observe that it implies that, for all halting programs of length $\le n$, the infinite part of each tape that comes after the $({\rm BB}_{U_p}(n))^{\rm th}$ bit is never read or modified by the TTM (moving there is by definition too long).

\begin{proposition}\label{ttm2}
A TTM with infinite tapes $(x,y)$ behaves with arbitrarily high probability exactly as if these infinite tapes were (extremely long but) finite:\\ $\forall \epsilon>0$, $\exists N$ such that \begin{equation}
P\left( {U_p}(x,y)=\left( U_p(x_{[\le N]},y_{[\le N]})~\,||\,~(x_{[> N]},y_{[> N]}) \right)  \right) \ge 1- \epsilon \,,
\end{equation}
where the subset notation is used to split $x=x_{[\le N]}\,||\,x_{[> N]}$ and $y=y_{[\le N]}\,||\,y_{[> N]}$.
\begin{proof}
Taking Eq.~\ref{mono} with $N \coloneqq{\rm BB}_{U_p}(N')$, with the consideration about busy beaver above (any machine that halts affects only a finite amount of tape).\qedhere
\end{proof}
\end{proposition}

Finally, Theorem~\ref{secondlaw} is obtained by combining Proposition~\ref{proposition6} and Proposition~\ref{ttm2}, with $\epsilon \rightarrow 0$:
\begin{equation}
\forall k>0,\, P\left(w_{\rm out} > f(\nu)+k\right) \le 2^{-k} \,,
\end{equation}
where $w_{\rm out}$ is the value of the \textbf{proof-of-work}.

\subsection{The exhaustive and sampled memory games}\label{memorygames}

We detail here in a game format a reduction that we later use in our security proofs. Our memory games involve an adversary against a verifier. The adversary sends, using a reversible channel \textsc{SWAP}, an exponentially long string to the verifier, but is also asked to try to keep a copy of it; the verifier then interrogates the adversary about either all of that string (in the \emph{exhaustive} variant), or about a random linear-size subset of it (in the \emph{sampled} variant); we show that the adversary has limited advantage in guessing as compared to a trivial strategy, unless they made an accurate copy of the whole string of exponential length --- a process that requires, in light of Landauer's principle, an exponential amount of either luck or free energy. We formalize this intuition, starting with the non-sampled version of the game.

\begin{definition}
The \emph{exhaustive $k\cdot2^{\nu} \choose k\cdot2^{\nu}$ memory game} is defined as follows for security parameters~$\nu$ and $k$:
\begin{enumerate}
\rm
\item The adversary isolates (by taking it from the environment of their lab for example) a system $X\in \mathcal{X}= \{0,1\}^{k\cdot2^{\nu}}$. All the rest of their available information is modelled as $E$.
\item The adversary (modelled as a TTM) makes some computation on the systems $X,E$.
\item Through a noiseless reversible channel (\emph{e.g.}, \textsc{SWAP}), the adversary sends $X$ to the verifier.
\item The verifier provides the adversary a blank tape of length $k\cdot2^{\nu}$, and asks the adversary to correctly print on it all of $X$.
\end{enumerate}
\end{definition}

\begin{proposition}\label{oneprop}
For any $2^\nu$-\bfe~adversary, the advantage at the exhaustive $k\cdot2^{\nu} \choose k\cdot2^{\nu}$ memory game, compared to a trivial coin-flip strategy, is bounded by
\begin{equation}
H_\infty(X|E)\ge (k-1)2^\nu\,.
\end{equation}

\begin{proof}
We reduce a violation of Theorem~1 (\emph{i.e.}, Landauer's principle) to a large advantage at the exhaustive $k\cdot2^{\nu} \choose k\cdot2^{\nu}$ memory game.
During the game, instead of sending $X$ to the verifier, the adversary deviates and XORs onto $X$ their best guess for $X$ given side information $E$. If the adversary guesses correctly, it turns $X$ into an all-$0$ string. This \textbf{proof-of-work} of length $k\cdot2^{\nu}$ violates Theorem~1 if it is created with probability higher than $2^{-(k-1)2^{\nu}}$; therefore, it does not.\qedhere
\end{proof}
\end{proposition}

The constraint also holds if the adversary is quizzed only on a random subset of positions.
\begin{definition}
The \emph{sampled $k\cdot2^{\nu} \choose t$ memory game} is defined as follows for free-energy bound $2^\nu$, security parameter $k$, and sample size $t$:
\begin{enumerate}
\rm
\item The adversary isolates (by taking it from the environment of their lab for example) a system $X\in \mathcal{X}= \{0,1\}^{k\cdot2^{\nu}}$. All the rest of their available information is modelled as $E$.
\item The adversary (modelled as a TTM) makes some computation on the systems $X,E$.
\item Through a noiseless reversible channel (\emph{e.g.}, \textsc{SWAP}), the adversary sends $X$ to the verifier.
\item The verifier chooses at random $t$ \textit{sample} positions $\subset\mathcal{X}$ and sends a description of these positions to the adversary, who must correctly guess $X_{[\textit{sample}]}$.
\end{enumerate}
\end{definition}

\begin{theorem}\label{twothm}
For any $2^\nu$-\bfe~adversary, the advantage at the sampled $k\cdot2^{\nu} \choose t$ memory game, compared to a trivial coin-flip strategy, is bounded, for all $\delta>0$, by
\begin{equation}
H_\infty^{\negl{t}}(X_{[\textit{sample}]}|E)\ge \frac{t\cdot (k-1)}{k} -t\cdot \delta \,.
\end{equation}

\begin{proof}
Lemma~6.2~in~\cite{vadhan2004constructing} states that, under random sampling, the min-entropy per bit is with high probability approximately conserved. In our case, this implies that, for all $\delta>0$,
\begin{equation}
H^{2^{-\Omega(t\delta^2 \log^2{\delta})}+2^{-\Omega(k 2^\nu \delta )}}_\infty(X_{[\textit{sample}]}|E) \ge \frac{t}{ k\cdot2^{\nu}} H_\infty(X|E)-t\cdot \delta\,,
\end{equation}
given which Theorem~\ref{twothm} follows from Proposition~\ref{oneprop}.\qedhere \end{proof}
\end{theorem}

\subsection{Universal hashing}\label{universalhash}
Universal hashing is useful for both privacy amplification and authentication.

\begin{definition}[2-universal hashing~\cite{carter1979universal,wegman1981new}]
Let $\mathcal{H}$ be a set of hash functions from $\{0,1\}^n \rightarrow \{0,1\}^m$. $\mathcal{H}$ is \emph{2-universal} if, given any distinct elements $x_1,x_2 \in\{0,1\}^n $ and any (not necessarily distinct) elements $y_1,y_2\in\{0,1\}^m$, then 

\begin{equation}
\# \{h\in \mathcal{H}| y_1{=}h(x_1)\land y_2{=}h(x_2)\}=\#\mathcal{H}/ 2^{2m} \,.
\end{equation}
\end{definition}

\begin{lemma}[Leftover hash lemma~\cite{bennett1988privacy,impagliazzo1989pseudo,haastad1993construction,bennett1995generalized}]\label{leftover}
Let $h: \mathcal{S} \otimes \mathcal{X}  \rightarrow \{0,1\}^m$ be a 2-universal hash function. If $H_\infty(X) \ge m + 2 \epsilon $, then
\begin{equation}
\delta \Big( (h(S,X),S) , U\otimes S \Big) \le  {2^{-\epsilon}}\,.
\end{equation}
$S$ is a short uniformly random seed and $X$ is the variable whose randomness is to be amplified. $U$ is the uniform distribution of appropriate dimension. The symbol $\otimes$ is used to represent the joint probability of independent distributions.
\end{lemma}

\section{Secret-Key Establishment}\label{secske}
Secret-key establishment (SKE) is a fundamental primitive for two-way secure communication because it allows for a perfectly secure one-time-pad encryption between Alice and Bob about which Eve knows nothing (otherwise the protocol aborts).
\subsection{Definitions (SKE)}

\begin{definition}
A secret-key-establishment scheme is \emph{sound} if, at the end the protocol, Alice and Bob possess the same key with overwhelming probability in the security parameter $\eta$:
\begin{equation}
P(K_A \neq K_B)\le {\negl{\eta}}\,.
\end{equation}
\end{definition}

\begin{definition}
A secret-key-establishment scheme is information-theoretically \emph{secure} (\emph{i.e.}, almost perfectly secret) if the key $K_B$ is uniformly random even given all of the adversary's side information $E$, except with probability at most negligible in the security parameter $\nu$:
\begin{equation}
\delta \Big( (K_B,E), U \otimes E \Big) \le {\negl{\nu}}\,.
\end{equation}

\end{definition}
In what follows, the variables $(A,B)\in (\mathcal{A},\mathcal{B})$ are strings from registers of length roughly $\bigO(\nu \log \nu)$, while
$(X,Y)\in (\mathcal{X},\mathcal{Y})$ denote strings from registers of length $\bigO(2^\nu)$.

\subsection{Protocol (SKE)}\label{protocolske}
\begin{theorem}
The following secret-key-establishment protocol is information-{\linebreak}theoretically sound and secure against any eavesdropper whose free energy is bounded by $2^\nu$. Alice and Bob need a quantity of free energy that is asymptotically $\bigO(\nu)$. \end{theorem}
Soundness is analyzed in Section~\ref{SKE sound}, and security in Section~\ref{skesecure}.

\begin{tcolorbox}[pad at break*=1mm,
  colback=gray!10!white,colframe=gray!75!black,title={Secret-key-establishment protocol:}]

\begin{enumerate}
	\item Alice starts \footnote{
The main parameters are\\
{- $\nu$, from the $2^\nu$ bound in free energy of Eve;\\
- $k$, which determines the tolerated error rate between Alice and Bob;\\
- $t$, the number of test bits to estimate the above error rate;\\
-~$s$, the length of the raw key (before processing).
}} with $X\in \mathcal{X}=\{0,1\}^{k\cdot 2^{\nu}}$ in a uniformly random state (extracted from the equidistributed environment of her lab). She draws uniformly at random a subset $\subset \{1,\dots,{k\cdot 2^{\nu}}\}$ of $s+t$ positions ${\it rawkey} $ and copies $({\it rawkey},X_{[{\it rawkey}]})\rightarrow A$ to her memory.
	\item Alice sends ${X}\rightarrow {Y} $ to Bob using a reversible channel ({\it e.g.}, a \textsc{SWAP} channel); it is possibly intercepted by Eve.\label{step2}
	\item Bob announces the receipt to Alice on an authenticated public channel. In case of no  receipt, they abort.\label{step3}
	\item Alice publishes the subset positions ${\it rawkey}$ on the (noiseless) authenticated public channel so that Bob can select $Y_{[{\it rawkey}]}\rightarrow B$. Alice and Bob draw a ${\it test}$ sub-subset of $t$ bits that they sacrifice to estimate the error rate $p_{\rm error}$ between $A$ and $B$.
	\item If the estimated $p_{\rm error}$ is too large, they abort. Otherwise, Alice and Bob apply information reconciliation (detailed in Section~\ref{SKE sound}) on the remaining $s$ bits $A_{[{\overline{\it test}}]}$ and $B_{[ {\overline{\it test}}]}$. 
	\item Alice and Bob apply privacy amplification (detailed in Section~\ref{skesecure}) and obtain a shared secret key of length $\approx ((k-1)/k- h_b(p_{\rm error}))\cdot s$.
\end{enumerate}
\end{tcolorbox}

$h_b(p)\coloneqq -p\log_2 p - (1-p) \log_2 (1-p)$ is the \emph{binary entropy}.

Note that for any fixed $p_{\rm error}$ (as long as it is not trivially $1/2$), Alice and Bob can choose a security parameter $k$ for which the protocol will be secure for that value of $p_{\rm error}$. That is unlike, for example, the BB84 quantum-key-distribution protocol, which only tolerates error rates less than $1/4$ (any more and Eve can intercept the whole quantum state).

\subsubsection{The intuition.}
Because she is $2^\nu$-bounded in free energy, Eve cannot copy to her memory the whole $k\cdot 2^{\nu}$-long string $Y$ that she sends to Bob, on which Bob will later base the raw key. Alice circumvents this limitation by already knowing the raw-key positions at the moment she sends $X$ ($X$ becomes, after Eve's potential tampering, $Y$) and thus need not store more than an asymptotically $\bigO\left(\nu\right)$-long segment of the $k\cdot 2^{\nu}$-long string. As in quantum key distribution, Eve can force the protocol to abort.

\subsection{Soundness analysis (SKE)}\label{SKE sound}
\subsubsection{Parameter estimation.}
We first estimate (using upper bounds) between Alice and Bob the global error rate $p_{\rm error}$ and the non-tested {\it rawkey} error rate $p_{\rm error}^{\overline{\it test}}$. The former quantity is important for the privacy amplification analyzed in Section~\ref{skesecure}, while the second is needed to analyze information reconciliation.

\begin{proposition}
Alice and Bob can accurately estimate the error rate $p_{\rm error} $ by sampling on the $t$ ${\it test}$ positions the error rate $p_{\rm error}^{\it test}$:
\begin{equation}
P\left( p_{\rm error}  \le p_{\rm error}^{\it test} +  \epsilon \right) \ge 1-{ e^{-2 \epsilon^2 t}}\,.\label{reshuffling}
\end{equation}
\begin{proof}
$p_{\rm error}^{\it test}$ is computed from the Hamming weight $\omega(\overline{{A_{[{\it test}]}\oplus B_{[{\it test}]}}})=t(1-p_{\rm error}^{\it test} )$. Chernoff's inequality bounds $p_{\rm error}$.\qedhere
\end{proof}
\end{proposition}

\begin{proposition}
Alice and Bob can accurately estimate $p_{\rm error}^{\overline{\it test}} $ from $p_{\rm error}^{{\it test}} $:
\begin{equation}
P\left( p_{\rm error}^{\overline{\it test}}  \le p_{\rm error}^{{\it test}} + \frac{s\cdot \epsilon}{s+t}\right) \ge 1-{ e^{-2 \epsilon^2 t}}\,.
\end{equation}
\begin{proof}
We insert $p_{\rm error}= (s\cdot p_{\rm error}^{\overline{\it test}} + t\cdot p_{\rm error}^{{\it test}})/(s+t)$ in Eq.~\ref{reshuffling} and isolate $p_{\rm error}^{\overline{\it test}} $.\qedhere
\end{proof}
\end{proposition}

\subsubsection{Information reconciliation (error correction).}\label{correction}

Once they have a good estimate of $p_{\rm error}^{\overline{{\it test}}}$, Alice and Bob achieve information reconciliation by applying error correction on that unused subset $\overline{{\it test}}$ of $s$ bits.

Note that it is important that the established key be based on Bob's string, rather than on Alice's, because the reasoning (see the security analysis in Section~\ref{skesecure}) using the sampled memory game only directly bounds from above the mutual information between Bob and Eve, not the one between Alice and Eve.
\begin{proposition}\label{errorprop}
For any non-trivial constant $p_{\rm error}^{\overline{{\it test}}}\neq 1/2$, Alice and Bob can transform the samples $A_{[{\overline{\it test}}]},B_{[ {\overline{\it test}}]}$ into the (non-necessarily secret) keys $K'_A,K'_B$ for which
\begin{equation}
P\left( K'_A = K'_B \right) \ge 1- {\negl{\eta}}\,.
\end{equation}
They can do so with $w\approx  h_b(p_{\rm error}^{\overline{{\it test}}})\cdot s$ (the exact value is given below) bits of authenticated public communication.
\end{proposition}

We present one standard construction to correct an arbitrary error rate on the~$s$ bits of \textit{rawkey} that were not used during the parameter-estimation phase.

\paragraph{Asymptotically optimal protocol for information reconciliation~\cite{brassard1993secret}:}
~

Let $w\coloneqq  \ceil{ s\cdot  h_b(p_{\rm error}^{\overline{{\it test}}} + \delta')+\eta }$;
\begin{enumerate}
\item Bob picks at random a hash function $h:\{0,1\}^s\rightarrow \{0,1\}^w$ from a 2-universal family $\mathcal{H}$ and computes $h(B_{[\overline{{\it test}]}})$.
\item Bob communicates $h$ and $h(B_{[\overline{{\it test}]}})$ to Alice, using the authenticated public channel.
\item Alice computes $\tilde{A}_{[\overline{{\it test}]}} \coloneqq \displaystyle \argmin_{x \in \{0,1\}^{\len{s}}} \left(\omega(x,A_{[\overline{{\it test}]}})| h(x){=}h(B_{[\overline{{\it test}]}})\right)$.
\end{enumerate}

Here, $\omega(\cdot,\cdot)$ is the Hamming distance;
$\delta'$ determines efficiency and $\eta$ is the security parameter.

\begin{proof}
We first count, in the uniform distribution, the smooth number of strings with length~$s$ that contains approximately $p_{\rm error}^{\overline{\it test}}$: Let $M\coloneqq\{x \in \{0,1\}^s\,|\, p_{\rm error}^{\overline{{\it test}}} - \delta' \le p_{\rm error}^{\overline{{\it test}}}(x)\le p_{\rm error}^{\overline{{\it test}}} + \delta'\}$; from the asymptotic equipartition property, we have $\forall \delta'>0$,
\begin{equation}
P\left( \#M\le 2^{s\cdot h_b(p_{\rm error}^{\overline{{\it test}}} + \delta')} \right) \ge 1-{2^{-{ \Theta(\eta)}}}\,.
\end{equation}

Because $\mathcal{H}$ is 2-universal, the probability of obtaining a correct hash from a non-correct candidate in $M$ is bounded by $2^{-w}$. By the union bound, the protocol is therefore sound except with probability at most $2^{-w}\cdot \#M$, which is $\negl{\eta}$.\qedhere
\end{proof}

While the above ideal information reconciliation protocol is optimal, it offers no (known) efficient way (in the computational complexity sense) for Alice to decode Bob's codeword. While we are in this work only concerned with thermodynamic (rather than computational) efficiency, we refer to \cite{brassard1993secret}, or to the theory of Shannon-optimal efficient algebraic codes, such as convoluted codes, for asymptotically ideal information-reconciliation protocols that are also computationally efficient.

\subsection{Security analysis (SKE)}\label{skesecure}
If the protocol does not abort, Eve has negligible information about the key $K_B$ at the end.
This security resides on the fact that even if Eve intercepts $X$ (which was sent from Alice to Bob) and replaces it with $Y$, she cannot keep roughly more than a fraction $1/k$ of the information about~$Y$. Thus, since the key is based on $Y$, Eve has limited knowledge about it.

Formally, this can be analyzed with the sampled $k\cdot 2^{\nu} \choose s$ memory game in Section~\ref{memorygames}. Theorem~\ref{twothm} thereat guarantees a good starting point --- Eve (who is $2^{\nu}$-$\bfe$) must have limited information about Bob's raw key of length $s$:
\begin{equation}
\forall \delta >0, \,H^{\negl{\nu}+\negl{s}}_\infty(Y_{[\overline{{\it test}}]}|E,{\it rawkey},\overline{{\it test}})=s \cdot \frac{k-1}{k}-s\cdot \delta\,.\end{equation}
The next step is to go from \emph{low} information to \emph{essentially no} information.

\subsubsection{Privacy amplification.}\label{privacy}
Privacy amplification turns a long string about which the adversary has potentially some knowledge into a shorter one about which the adversary has essentially none.

In secret-key establishment, Eve's partial information can come from eavesdropping (and as shown, this quantity is roughly a fraction $1/k$) or from the public information leaked by the information reconciliation protocol, which is easily characterized.

Privacy amplification can be realized in an information-theoretically secure manner with 2-universal hashing (see Section~\ref{universalhash}).

\begin{proposition}
After privacy amplification, $K_B$ is approximately of length $\approx ((k-1)/k- h_b(p_{\rm error}))\cdot s$, and Eve has essentially no knowledge about it.
\begin{proof}
Let $w$ quantify the number of bits about $B_{[\overline{{\it test}]}}$ exchanged publicly during the information-reconciliation (IR) protocol.
We note that $H_\infty(K_B|E^{\rm pre IR})\le H_\infty(K_B|E^{\rm post IR})-w$,
hence
\begin{equation}
\forall \delta>0, \,H^{\negl{\nu}+\negl{s}}_\infty(K_B|E^{\rm post IR})=s \cdot \frac{k-1}{k}-s\cdot \delta-w \,.
\end{equation}
Therefore, taking $m\coloneqq s \cdot \frac{k-1}{k}-s\cdot \delta-w-\epsilon $ guarantees after hashing ($\epsilon$ is the security parameter for the Leftover hash lemma; see Section~\ref{universalhash}) information-theoretic security on those remaining $m$ bits.\qedhere
\end{proof}
\end{proposition}

Note that for any fixed $p_{\rm error}$, the parameters $s$ and $k$ can be selected as to make $m$ a positive quantity when the protocol does not abort (as a result of too many errors). Also note that the parameters $\nu$ and $s$ must not be too small.

\newpage

\section{1-out-of-2 Oblivious Transfer}\label{secot}
Oblivious transfer (OT) is a cryptographic primitive that is universal for two-party computation~\cite{rabin1981exchange,kilian1988founding}. It comes in many flavours, but they are all equivalent~\cite{crepeau1987equivalence}. We concern ourselves with 1-out-of-2 OT (or 1--2 OT). Informally: Alice sends two envelopes to Bob; Bob can open one to read the message in it, but he cannot open both; Alice cannot know which message Bob read.

\subsection{Definitions (OT)}
\begin{definition}
A 1--2 OT protocol is perfectly \emph{sound} if, when Alice and Bob are honest, the message $B(i)$ received by Bob is with certainty the message $m_i$ sent by Alice, for his choice of $i\in\{0,1\}$:
\begin{equation}
P\left(B (i)=m_i \right)=1\,.
\end{equation}
\end{definition}

\begin{definition}
A 1--2 OT protocol is information-theoretically \emph{secure-for-Alice} if Bob cannot learn something non-negligible about both of Alice's messages simultaneously: For any $2^{\nu}$-$\bfe$ Bob,\begin{equation}
\exists j \mbox{~s.t.~}\delta \Big( (m_{j},E_B), (U \otimes E_B)\Big) \le \negl{\eta}   \,.
\end{equation}
\end{definition}
$E_B$ denotes all of (a potentially malicious) Bob's side information. And similarly for $E_A$ in regards to Alice.

\begin{definition}
A 1--2 OT protocol is information-theoretically \emph{secure-for-Bob} if Alice cannot learn anything non-negligible about Bob's choice $i$: For any $2^{\nu}$-$\bfe$ Alice,
\begin{equation}
\delta\Big( (i,E_A), U \otimes E_A \Big) \le {\negl{\eta}}\,.
\end{equation}
\end{definition}
An OT protocol is information-theoretically secure when it is information-{\linebreak}theoretically secure for \emph{both} Alice and Bob.

\subsection{Protocol (OT)}\label{protocolot}

\begin{theorem}
The following 1--2 OT protocol is perfectly sound and information-theoretically secure against $2^\nu$-$\bfe$ adversaries. The free-energy requirement of the honest players is asymptotically $\bigO(\nu)$. \end{theorem}
The perfect soundness is straightforward. Security is analyzed in Section~\ref{OT security}.

\begin{tcolorbox}[pad at break*=1mm,
  colback=gray!10!white,colframe=gray!75!black,title={1--2 oblivious-transfer protocol:}]

(The variable $\eta$ is a security parameter.)

\begin{enumerate}
\item Alice chooses messages $m_0$ and $m_1$ of length $n$.
\item Alice starts with the exponentially long bit strings $X^{(0)},X^{(1)} \in \mathcal{X}=\{0,1\}^{4\cdot 2^{\nu}}$ in uniformly random states. She picks a random subset $\subset \{1,\dots,4 \cdot 2^{\nu}\}$ of $n+\eta$ positions ${\it raw}$ and stores $({\it raw},X^{(0)}_{[\it raw]}, X^{(1)}_{[\it raw]})$ in her memory.
\item Alice sends $(X^{(0)},X^{(1)})$ to Bob using the reversible channel \textsc{SWAP}.
\item Bob chooses $i\in \{0,1\}$ and computes reversibly $(X^{(0)},X^{(1)}) \rightarrow (X^{(i)},X^{(0\oplus1)})$, where we define $X^{(0\oplus 1)}\coloneqq X^{(0)}\oplus X^{(1)}$. Then, Bob keeps $X^{(i)}$ and sends back $X^{(0\oplus1)}$ reversibly to Alice using \textsc{SWAP}.
\item Alice receives $\tilde{X}^{(0\oplus1)}$ and checks whether $\tilde{X}^{(0\oplus1)}_{[\it raw]}{=}X^{(0\oplus1)}_{[\it raw]}$. If they differ, Alice aborts.\label{perfect match step}
\item Alice chooses at random a 2-universal hash function $h: \{0,1\}^{n+\eta} \rightarrow \{0,1\}^{n}$ and communicates $h,{\it raw},m_0 \oplus h(X^{(0)}_{[\it raw]}),m_1 \oplus h(X^{(1)}_{[\it raw]})$ to Bob.
\item Bob computes the hash $h(X^{(i)}_{[\it raw]})$ and recovers $m_i$. 
\end{enumerate}
\end{tcolorbox}

\subsubsection{The intuition.}
In addition to the previously exploited \emph{impossibility to copy} exponential quantities of information without using corresponding quantities of free energy or violating Landauer's principle, the oblivious-transfer protocol makes use of another key feature of \emph{reversible computing}: As long as Bob is in possession of $X^{(0\oplus 1)}\coloneqq X^{(0)}\oplus X^{(1)}$, the maximally random variables $X^{(0)}$ and $X^{(1)}$ have conditionally exactly the \emph{same} information content; but once $X^{(0\oplus 1)}$ is returned to Alice, $X^{(0)}$ and $X^{(1)}$ revert to being \emph{uncorrelated}. In other words, although sending $X^{(0\oplus 1)}$ back to Alice forces Bob to \emph{forget} information about the couple $X^{(0)},X^{(1)}$ (enabling 1-out-of-2 transfer), it does not uniquely specify \emph{which} information he forgot (Alice remains oblivious).

\subsection{Security analysis (OT)}\label{OT security}
\subsubsection{Security for Bob.}
From Alice's point of view, Bob's behaviour (\emph{i.e.}, sending ${X}^{(\rm 0\oplus 1)}$ back to Alice) is identical whether he chooses message $i{=}0$ or message $i{=}1$; the scheme is therefore perfectly secure for Bob.
\subsubsection{Security for Alice.}
We prove that a malicious Bob cannot learn anything non-negligible about a second message as soon as he learns something non-negligible about a first message.
\begin{proof}

We pose without a loss of generality that $\omega$ is the event corresponding to ``Bob learns something non-negligible about $m_0$.'' Because he is $2^\nu$-bounded in free energy, a malicious Bob's success at the sampled $4 \cdot 2^{\nu} \choose n+\eta $ memory game (on state $\tilde{X}^{(\rm 0\oplus 1)}$ and sample {\it raw}) is bounded by Theorem~\ref{twothm}:
\begin{equation}
\forall \delta>0,\, H_\infty^{\negl{\nu}+\negl{\eta}}(\tilde{X}^{(0\oplus1)}_{[\it raw]}|E_B,\omega)\ge (n+\eta)/2 - (n+\eta)\cdot \delta\,.\label{contra1}
\end{equation}
By subadditivity, we have
\begin{align}
&H_\infty^{\negl{\nu}+\negl{\eta}}(\tilde{X}^{(0\oplus1)}_{[\it raw]}|E_B,\omega) \\&\le H_\infty^{\negl{\nu}+\negl{\eta}}({X}^{(0)}_{[\it raw]},{X}^{(1)}_{[\it raw]}|E_B,\omega)\\ &\le H_\infty^{\negl{\nu}+\negl{\eta}}({X}^{(0)}_{[\it raw]}|E_B,\omega)+H_\infty^{\negl{\nu}+\negl{\eta}}({X}^{(1)}_{[\it raw]}|E_B,\omega) \,.
\end{align}
We apply the Leftover hash lemma (Lemma~\ref{leftover}) with $\epsilon \coloneqq \eta/12-3n/8$. The two privacy-amplification steps succeed (except by the union bound with probability $\negl{\nu}+\negl{\eta}$) if, respectively,
\begin{align}
&H_\infty^{\negl{\nu}+\negl{\eta}}({X}^{(0)}_{[\it raw]}|E_B,\omega)  \ge n/4+\eta/6\,,\\
&H_\infty^{\negl{\nu}+\negl{\eta}}({X}^{(1)}_{[\it raw]}|E_B,\omega)  \ge n/4+\eta/6 \,.
\end{align}
We assume by contradiction that they are both unsuccessful with non-negligible probability. It implies
\begin{equation}
H_\infty^{\negl{\nu}+\negl{\eta}}(\tilde{X}^{(0\oplus1)}_{[\it raw]}|E_B,\omega) < n/2 + \eta/3\,,
\end{equation}
which contradicts Eq.~\ref{contra1} for small $\delta \le \eta/(6(n+\eta))$.\qedhere
\end{proof}

\section{From classical adversaries to quantum adversaries}\label{toquantum}

Up to here, the notion of information that has been used --- in the protocols for secret-key establishment and oblivious transfer, as well as in their analyses --- is purely \emph{classical}. But as scrutinised by thorough experiments~(notably, the extensive serie of Bell experiments~\cite{freedman1972experimental,aspect1982experimental,hensen2015loophole,giustina2015significant,shalm2015strong}), nature is \emph{quantum-physical}.
The aim of this section is to bring our work one step closer to the quantum realm. Namely, we investigate whether our (classical\footnote{All classical operations can be viewed as quantum operations restricted to diagonal density matrices.}) protocols are secure against quantum adversaries. We find that our SKE protocol (Section~\ref{protocolske}) is secure against a quantum Eve \emph{as it is}. On the other hand, to retain security against a malicious quantum Alice, our OT protocol (Section~\ref{protocolot}) has to be slightly updated --- the patched protocol presented below in Section~\ref{patchedprotocol} is quantum-safe but remains classical for honest players. Our work's conclusion, therefore, fully extends to the \emph{quantum} world of Maxwell demons (given arbitrarily large but random environments): It is~--- on paper ---~information-theoretically cryptographically friendly.

\subsection{The setting made quantum}
Our model described in Section~\ref{sectionmodel} is based on Alice, Bob, and Eve being classical computers with thermodynamical restrictions (we call them Thermodynamical Turing Machines) interacting through classical channels (a standard authenticated channel and a \textsc{SWAP} channel).

In a quantum setting, Alice, Bob, and Eve are upgraded to universal quantum computers~\cite{deutsch1985quantum} and their communication channels can carry states in quantum superposition. A quantum computer cannot compute more than a classical computer could (given exponential computational time, a classical computer can simulate a quantum computer). Quantum computing cannot either be used to evade Landauer's principle~\cite{faist2015minimal}. As such, once all elements are properly defined, a quantum version of our Theorem~\ref{thm:violation} holds.

\begin{proposition}[Thm.~\ref{thm:violation} in the quantum realm (sketch)]\label{quantumsecondlaw}
For all $k>0$, a player modelled by a quantum computer with a bound $f(\nu)$ in free energy cannot erase more than $f(\nu)+k$ initially completely mixed qubits, except with probability $2^{-k}$.
\end{proposition}

The ability to send and receive quantum states does enable new possibilities for both honest and malicious agents --- we investigate next how this affects the security of our previous SKE and OT protocols.

\subsection{The quantum exhaustive and sampled memory games}
We extend the proof method developed in Section~\ref{memorygames} to the quantum world.

First, the bound on the success of an adversary at the exhaustive $k\cdot2^{\nu} \choose k\cdot2^{\nu}$ memory game (Proposition~\ref{oneprop}) is unaffected by the transition from classical to quantum information.

\begin{proposition}[Prop.~\ref{oneprop} with quantum side-information]\label{onepropQ}
For any quantum adversary with a bound $2^\nu$ in free energy, the advantage at the exhaustive $k\cdot2^{\nu} \choose k\cdot2^{\nu}$ memory game, compared to a trivial coin-flip strategy, is bounded by
\begin{equation}
H_\infty(X|E)\ge (k-1)2^\nu\,.
\end{equation}
\begin{proof}
$X$ is here still classical, but $E$ represents side information that is possibly quantum. Since the operational meaning of conditional min-entropy is the same whether the side information is quantum or not~\cite{konig2009operational}, the argument presented in Section~\ref{memorygames} is unchanged.\qedhere\end{proof}
\end{proposition}

The next step is to sample from $X$ (Theorem~\ref{twothm}).

\begin{proposition}[Thm.~\ref{twothm} with quantum side-information]\label{twothmQ}
For any quantum adversary with a bound $2^\nu$ in free energy, the advantage at the sampled $k\cdot2^{\nu} \choose t$ memory game, compared to a trivial coin-flip strategy, is bounded, for all $\delta>0$, by
\begin{equation}
H_\infty^{\negl{t}}(X_{[\textit{sample}]}|E)\ge \frac{t\cdot (k-1)}{k} -t\cdot \delta \,.
\end{equation}
\begin{proof}
The result by Vadhan~\cite{vadhan2004constructing} that we used in the classical case has been generalized in presence of quantum side information by K\"onig and Renner in~\cite{konig2011sampling}. Apart from the exact parameter values hidden behind $\negl{t}$, our proof is, hence, unchanged by the addition of quantum side information.
\end{proof}
\end{proposition}

\subsection{The classical SKE protocol is already quantum-resistant}
The information-theoretical security of the SKE protocol from Section~\ref{secske} depends uniquely on the one of privacy amplification and on Theorem~\ref{twothm}.

Since in presence of quantum side information, universal-2 hashing (Lemma~\ref{leftover}) remains a universally composably secure way of achieving privacy amplification~ \cite{renner2005universally,tomamichel2011leftover}, and that, as we just argued, so is the case of Theorem~\ref{twothm}, the SKE scheme presented in Section~\ref{protocolske} is secure against quantum adversaries.

Fundamentally different from standard quantum key distribution, the result is nevertheless an information-theoretically secure key distribution scheme for a quantum world in which entropy is exponentially cheaper than free energy.

\subsection{A quantum-resistance patch for the OT protocol}\label{patchedprotocol}
\begin{figure}[h!]
\begin{center}
\hspace{-1em}\includegraphics[width=0.55\textwidth]{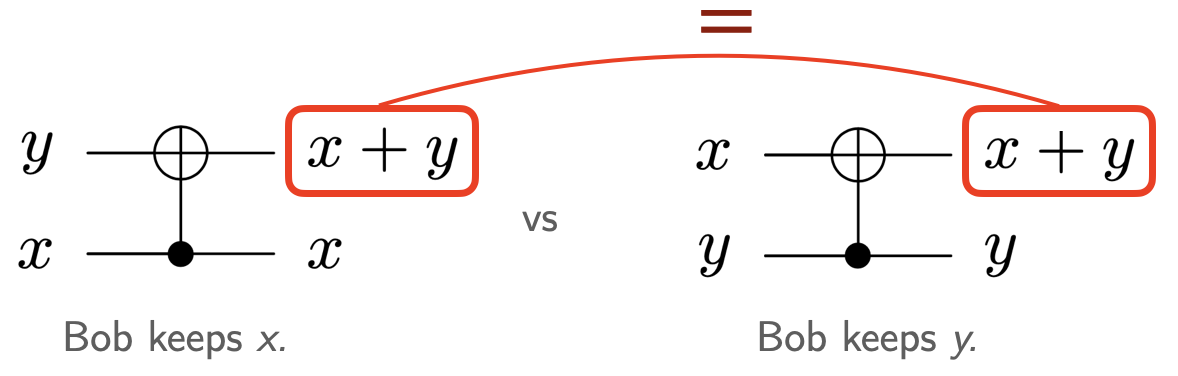}
\caption{If Bob receives a classical state, the top state, $x+y$, that he will return to Alice during the OT protocol will be the same no matter whether he chooses to decrypt the first (left) or second message (right).}
\label{figurexy}
\end{center}
\end{figure}

\begin{figure}[h!]
\begin{center}
\hspace{-2em}\includegraphics[width=0.9\textwidth]{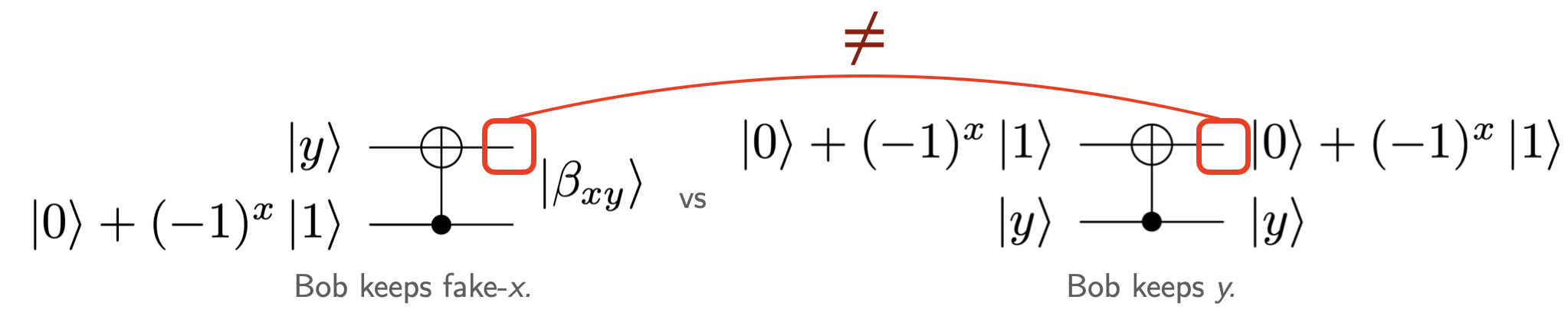}
\caption{A malicious Alice can send to Bob one of the quantum states in the Hadamard basis. In that case, the upper state sent back to Alice by an honest Bob will be $\ket{+}$ or $\ket{-}$ if he wants to keep the first message, but half of one of the four Bell states $\{\ket{\beta_{xy}}\}_{xy}$ if he wants to keep the second message. Since Alice can distinguish between those two cases, the OT scheme is not secure for Bob. Below, we explain how Bob can prevent this quantum attack.}
\label{figurexpasy}
\end{center}
\end{figure}

Given that the above SKE protocol is quantum-resistant, and that the same argument applies to the security-for-Alice part of our oblivious-transfer protocol, it would be natural for our previously detailed scheme to be also quantum-resistant. But it is not: The security-for-Bob, which is trivial in the classical case (because $x+y=y+x$, see Fig.~\ref{figurexy}), can be broken by a malicious quantum Alice. The reason is that if Alice acts maliciously and sends the superposed quantum states $X^{(0)}=H\ket{x}$ and $Y^{(0)}=\ket{y}$ to Bob (for some random $x$ and $y$), she can discriminate between the state sent back by Bob when he does $H\ket{x} \overset{\rm CNOT}{\longrightarrow} \ket{y} $ (to keep $X^{(0)}$) compared to when he does $\ket{y} \overset{\rm CNOT}{\longrightarrow} H\ket{x}$ (to keep $Y^{(0)}$). This attack is illustrated in Fig.~\ref{figurexpasy}.


But there is a simple patch for this attack, or, in fact, for all quantum attacks by a malicious Alice. Alice's extra power comes from the fact she can send states in superposition, but Bob can in return pre\"emptively ``classicize'' the possibly quantum states ${X^{(0)}}$ and ${X^{(1)}}$ by CNOT-ing each bit to a different bit of the totally mixed environments $\pi_0$ and $\pi_1$. Given control of a large enough environment (of dimension $2^{\len{X^{(0)}}+\len{X^{(1)}}}$), Bob can do so at no free energy cost. The resulting state, when traced over that environment, is then undistinguishable from a (possibly noisy) state sent by a malicious-but-classical Alice. Even if misbehaviour from Alice's part might affect the protocol's correctness (which is allowed for a malicious Alice), it leaves the perfect security intact: a quantum Alice can still not gain any information about Bob's choice.

\begin{tcolorbox}[pad at break*=1mm,
  colback=gray!10!white,colframe=gray!75!black,title={Quantum-safe 1--2 oblivious-transfer protocol}]

Steps~1--3~and~5--7 are the same as in the previous classical protocol. Step~4 is changed to

\begin{enumerate} \setcounter{enumi}{3}
\item[4'.] Bob chooses $i\in \{0,1\}$ and computes reversibly
\begin{align*} (X^{(0)},X^{(1)},\pi_0,\pi_1) \rightarrow (X^{(i)},X^{(0\oplus1)},\pi_0 \oplus{X^{(1)}},\pi_1 \oplus{X^{(2)}})\,,
\end{align*} where $\pi_0$ and $\pi_1$ are completely mixed states of appropriate size taken from Bob's environment, and where we define $X^{(0\oplus 1)}\coloneqq X^{(0)}\oplus X^{(1)}$. Then, Bob keeps everything but $X^{(0\oplus1)}$, which he sends back (thermodynamically reversibly) to Alice using \textsc{SWAP}.
\end{enumerate}
\end{tcolorbox}

The above step reduces the security for Bob in the quantum case to the one of the classical case. The updated protocol does not require the honest players to make any quantum operations \emph{per se}.

\section{Concluding remarks}
We propose a \emph{free-energy-bounded} model of cryptography, in which we have derived information-theoretically secure protocols for secret-key establishment and oblivious transfer.

Even if the rationale behind its security is totally different: Our secret-key-establishment protocol is similar to standard quantum key distribution. Our oblivious-transfer protocol, on the other hand, is novel in itself: The mechanism that allows Alice to check that Bob honestly forgets information is proper to reversible computing.

Our schemes are not practical at this point: Current technology is still far from computing with memories that are large enough for Landauer's principle to become the main obstacle (it is worth noting that Boltzmann's constant, which we have in this work conveniently set to $k_{\rm B}:=1/{\rm T}$, is in fact $\approx 1.38\cdot 10^{-23} {\rm J K}^{-1}$); and whereas no laws of physics forbid it, implementing reversible computation on such states is for now science fiction. Our result is rather to be seen as part of the quest of distinguishing what physical phenomena allow for realizing cryptographic functionalities in principle, and which do not. In this spirit, our protocols add another element to the longer and longer list of physical laws from which cryptographic security \emph{can} directly be derived:
We can now claim that information-theoretic key agreement
is theoretically possible as soon as one of the fundamental limits conjectured by \emph{either} quantum theory \emph{or} special relativity \emph{or} \emph{the second law of thermodynamics} is correct.
Concerning the novel appearance of a thermodynamic law in this list, we remark first that according to \emph{Albert Einstein},
thermodynamics is the only physical theory that will survive future development in Physics. Second, the second law is
rather pessimistic in nature, and to see it being linked to a constructive application is refreshing.
We are, in fact, not aware of many uses, besides our protocols, of the law. In summary, we can say, somewhat ironically: \emph{One small step for cryptography --- one giant leap for the second law.}

\subsubsection*{Acknowledgement}

We thank Charles Alexandre B\'edard and Cecilia Boschini for helpful comments about earlier version of this work.
We also thank Renato Renner and two anonymous referees for some interesting remarks regarding the physicality of the model.
This research was supported by the Swiss National Science Foundation (SNF).

\bibliographystyle{alpha}
\bibliography{../ddd_all.bib}

\end{document}
\documentclass[]{llncs}
\let\proof\relax
\let\endproof\relax
\usepackage{amsmath,amssymb,amsthm}
\usepackage{tikz}
\usepackage[utf8]{inputenc}
\usepackage[T1]{fontenc}
\usepackage{parskip} \usepackage{physics}
\usepackage{mathtools}
\usepackage{enumitem}
\usepackage[most]{tcolorbox}
\usepackage{hyperref}
\usepackage{qcircuit}
\usepackage{float}

\newif\ifshowdiscussion
\showdiscussiontrue \newcommand{\discussion}[1]{\ifshowdiscussion\textcolor{red}{~{#1}}\fi}
\newcommand{\discussionb}[1]{\ifshowdiscussion\textcolor{blue}{~{#1}}\fi}
\newtheorem{assum}{Assumption}
\newtheorem*{landauer*}{Landauer's principle}
\newtheorem*{szil\'ard*}{Szil\'ard's engine}
\newtheorem*{rev*}{Reversible computation}
\newtheorem*{finetti*}{Finite de Finetti's Theorem}

\let\definition\relax
\spnewtheorem{definition}{Definition}{\bfseries}{}

\newcommand{\bigO}{\mathcal{O}}

\newcommand{\len}[1]{{\bf len}(#1)}
\newcommand{\negl}[1]{{\bf negl}(#1)}
\newcommand{\poly}[1]{{\bf poly}(#1)}
\newcommand{\spoly}[1]{{\bf superpoly}(#1)}
\newcommand{\nnegl}[1]{{ \textbf{non-negl}}(#1)}
\newcommand{\expp}[0]{{\bf exp}}
\newcommand{\ewp}[1]{\tag*{\scalebox{1}{\qquad\rm$[ \mbox{e.w.p.}\le #1]$}$\,.$}}
\newcommand{\ewpv}[1]{\tag*{\scalebox{1}{\qquad\rm$[ \mbox{e.w.p.}\le #1]$}$\,,$}}
\newcommand{\wnnp}[1]{\tag*{\scalebox{1}{\qquad\rm$[ \mbox{w.p.~} \nnegl{#1}]$}$\,.$}}
\newcommand{\wnnpv}[1]{\tag*{\scalebox{1}{\qquad\rm$[ \mbox{w.p.~} \nnegl{#1}]$}$\,,$}}
\DeclarePairedDelimiter{\ceil}{\lceil}{\rceil}
\DeclareMathOperator*{\argmin}{\arg\!\min}

\usepackage{epigraph}

\setlength\epigraphwidth{8cm}
\setlength\epigraphrule{0pt}

\usepackage{etoolbox}

\makeatletter
\patchcmd{\epigraph}{\@epitext{#1}}{\itshape\@epitext{#1}}{}{}
\makeatother

\title{Key Agreement and Oblivious Transfer\\ from Free-Energy Limitations}
\author{Xavier Coiteux-Roy  \and Stefan Wolf}
\institute{Universit\`a della Svizzera italiana, Lugano, Switzerland.\\
\email{\{xavier.coiteux.roy,stefan.wolf\}@usi.ch}}
\date{\today}  

\begin{document}

\maketitle

\noindent
\makebox[\linewidth]{\small \today}

\begin{minipage}{1.01\linewidth}
~
\begin{abstract}
We propose one of the very few \emph{constructive} consequences of the second law of thermodynamics. More specifically, we present protocols for secret-key establishment and multiparty computation the security of which is based fundamentally on Landauer's principle. The latter states that the erasure cost of each bit of information is at least $k_{\rm B} \rm{T} \ln 2$ (where $k_{\rm B}$ is Boltzmann's constant and ${\rm T}$ is the absolute temperature of the environment). Albeit impractical, our protocols explore the limits of reversible computation, and the only assumption about the adversary is her inability to access a quantity of free energy that is exponential in the one of the honest participants. Our results generalize to the quantum realm.
\end{abstract}
\end{minipage}
\vspace{0.5cm}
\pagestyle{plain} 

{\bf Keywords:} Reversible computation, quantum information, information-\\theoretic security, key establishment, oblivious transfer.

\section{Introduction}
\subsection{Motivation}
In the past decades, several attempts were made to achieve cryptographic security from physical properties of communication channels: Most prominently, of course, \emph{quantum cryptography}~\cite{bb84,ekert1991quantum}; other systems made use of noise in communication channels~\cite{wyner1975wire} or bounds on the memory space accessible by an adversary~\cite{maurer1992conditionally}. These schemes have in common that no limit is assumed on the opponent's computational power: They are \emph{information-theoretically secure}. 

Our schemes for achieving confidentiality (key agreement or, more precisely, \emph{key expansion}) as well as secure co\"operation (multiparty computation, \emph{i.e.}, \emph{oblivious transfer}) rely solely on a bound on the accessible \emph{free energy}\footnote{Free energy is ``free'' in the sense that it can be used to do work --- it is not ``entrapped'' in a system.} of an adversary. More specifically, we propose schemes the security of which follows from \emph{Landauer's principle}, which is a quantification of \emph{the second law of thermodynamics}: \emph{In a closed system, ``entropy'' does not decrease} (roughly speaking). 

\emph{Landauer's principle} states that the \emph{erasure of information} unavoidably costs free energy, the amount of which is proportional to the length of the string to be erased. On the ``positive'' side, the \emph{converse} of the principle states that the all-$0$ string of length $N$ has a free-energy value proportional to $N$. More precisely, the erasure cost and work value are both quantified by $k_{\rm B} {\rm T}\ln 2\cdot N$, where $k_{\rm B}$ is \emph{Boltzmann's constant} (in some sense the nexus between the micro- and macroscopic realms), and ${\rm T}$ is the absolute temperature of the environmental heat bath. 

Our result can be seen as one episode in a series of results suggesting information-theoretic security to be, in principle, achievable under the assumption that \emph{at least one in a list of physical theories}, such as quantum mechanics or special relativity, \emph{is accurate}: We add to this list the second law of thermodynamics~--- to which not much glamour has been attached before. 

\subsection{Contributions}
We base the ``free-energy-bounded model'' of information-theoretic cryptography upon the observation that the second law of thermodynamics has a cryptographically useful corollary: ``Copying information has a fundamental cost in free energy.'' Bounding the free energy of an adversary forces them into picking parsimoniously what to copy, and that can be exploited in a reversible-computing context to ensure information-theoretic security. Our secret-key establishment protocol demonstrates how bounds in free energy can lead to cryptographic mechanisms similar to the ones used in quantum-key distribution and in the bounded-storage model, while our oblivious-transfer protocol exemplifies the novelty of our model.

This is an overview of our article: In Section~\ref{stateoftheart}, we review the subjects of information-theoretic cryptography and of reversible computing. In Section~\ref{sectionmodel}, we introduce, based on reversible computing, a novel model of computation and interaction that captures the consumption and the production of free energy in Turing machines. In Section~\ref{secproof}, we establish some prerequisites: we prove a version of Landauer's principle in our framework, and construct a game that is basically equivalent to a thermodynamical ``almost-no-cloning theorem,'' which we later use in our security proofs. In Sections~\ref{secske}~and~\ref{secot}, we offer protocols for \emph{secret-key establishment} and \emph{oblivious transfer}, respectively; their information-theoretical security is based fundamentally on Landauer's principle. It is assured against adversaries whose bound in free energy is exponential compared to the one of the honest players. While the present work focuses on classical information, we sketch in Section~\ref{toquantum} how all our results generalize in presence of quantum adversaries.

\section{State of the Art}\label{stateoftheart}
\subsection{Information-theoretic cryptography from physical assumptions}

In parallel to the development of computationally secure
cryptography~--- and somewhat in its shadow~---, attempts were made to obtain in a provable fashion stronger, \emph{information-theoretic
   security},
based not on the hardness of obtaining the (uniquely
determined) message in question, but on the sheer lack of information. 
Hereby, the need for somehow 
``circumventing'' Shannon's pessimistic theorem of perfect secrecy is met by some sort of
\emph{physical 
limitation}. The latter can come in the form of simple noise in a
communication channel, a limitation on accessible memory, the
uncertainty principle of quantum theory, or
the 
non-signalling postulate of special relativity.

The first system of the kind, radically improving on the perfectly
secret
yet impractical \emph{one-time pad}, has been 
\emph{Aaron Wyner}'s wiretap channel~\cite{wyner1975wire}: 
Here, information-theoretic secret-key establishment becomes possible ---
under the assumption,
however, that the legitimate parties already start with an advantage, 
more specifically, that the adversary only has access to a
non-trivially
degraded version of the recipient's pieces of information. 
A \emph{broadcast scenario} was proposed by \emph{Csisz\'ar} and
\emph{K\"orner}~\cite{csiszar1978broadcast}~--- where, again, an initial advantage 
in terms of information proximity or information quality was required by the 
legitimate partners \emph{versus} the opponent. A~breakthrough 
was marked by the work of \emph{Maurer}~\cite{maurer1993secret}, who showed that the need
for such an initial advantage on the information level can be replaced
by \emph{interactivity} of communication: Maurer, in addition, 
conceptually simplified and generalized the model by separating 
the noisily correlated data generation from public yet
authenticated
communication, the latter being considered to be for free. 
The model shares its communication setting with both  
\emph{public-key} as well as \emph{quantum cryptography}. 
Maurer and Wolf~\cite{maurer1996towards} have shown that in the 
case of independent-channel access to a binary source, key agreement
is in fact possible in principle in \emph{all} non-trivial cases,
\emph{i.e.}, even when Eve starts with a massive initial advantage 
in information quality. 

In the same model, it has also been shown that \emph{multiparty
  computation} becomes possible, namely \emph{bit commitment}
and (the universal primitive of) \emph{oblivious transfer}~\cite{crepeau1988achieving,crepeau1997efficient}.
More generally, oblivious transfer has also been achieved from {\em
  unfair} 
noisy channels, where the error behaviour is prone to be influenced in
one way or another by the involved, distrusting parties willing to co\"operate. 

The \emph{public-randomizer model} by Maurer~\cite{maurer1992conditionally}
has generally been recognized as the birth of the idea of
``memory-bounded models,'' based on the fact that the \emph{memory} an 
opponent or cheater (depending on the context) can access is
limited. Specifically, Maurer assumes the wire-tapper can obtain a
certain \emph{fraction} of the physical bits. This was generalized to arbitrary \emph{types} of information
by \emph{Dziembowski} and Maurer~\cite{dziembowski2002tight}.
Analogously, also \emph{oblivious transfer} has been shown achievable 
with a memory-bounded receiver~\cite{cachin1998oblivious,ding2004constant}.
The main limitation to the memory-bounded model, for both secret-key establishment and multiparty computation, is that the memory advantage of the honest participants over the adversaries is at most quadratic~\cite{dziembowski2004generating}.

The idea to use \emph{quantum physics} for cryptographic ends dates
back 
to \emph{Wiesner}, who, for instance, proposed to use the uncertainty
principle 
to realize unforgeable banknotes. His ``conjugate coding''~\cite{wiesner1983conjugate} resembles
oblivious 
transfer; the latter~--- even bit commitment, actually~--- we know now to be unachievable from quantum
physics only~\cite{mayers1997unconditionally,lo1998quantum}. A breakthrough has been the now famous ``BB84''
protocol 
for key agreement by communication through a channel allowing for 
transmitting quantum bits, such as an optic fibre, plus a public yet
authenticated classical channel~\cite{bb84}. 

A combination of the ideas described is the ``bounded quantum-storage
model'' \cite{damgaard2008cryptography}: 
Whereas no quantum memory is needed at all for the honest players, a
successful adversary can be shown to need more than $n/2$ of the
communicated quantum bits.
The framework has been unified and generalized to the ``noisy'' model
by \emph{K\"onig}, \emph{Wehner}, and \emph{Wullschleger}~\cite{konig2012unconditional}.

Very influential has been a proof-of-principle result
by \emph{Barrett, Hardy, and {\linebreak}Kent}~\cite{barrett2005no}: The
security in key agreement that stems from witnessing quantum correlations can be established regardless of the validity of quantum theory, only from the postulate of special relativity that there is \emph{no superluminal signalling}. The authors combined \emph{Ekert}'s~\cite{ekert1991quantum}
idea to obtain secrecy from proximity to a pure state, guaranteed by 
\emph{close-to-maximal violation of a ``Bell inequality,''} with the 
role this same ``nonlocality'' plays in the argument that the
outcomes
of quantum measurements are, in fact, random and not predetermined: 
In the end, reasoning results that are totally \emph{independent} of the completeness of quantum 
theory. Later, efficient realizations of the paradigm were
presented~\cite{hanggi2010efficient,masanes2011secure}. 
Conceptually, an interesting resulting statement is that
information-theoretic 
key agreement is possible if \emph{either quantum mechanics OR
  relativity theory} are complete and accurate
``descriptions of nature.''  
Another point of interest is that trust in the manufacturer is not
even required: ``device independence''~\cite{PhysRevLett.113.140501}.

\emph{Kent} also demonstrated that bit commitment can be information-theoretically secure thanks to special relativity alone~\cite{kent1999unconditionally}. On the other hand, oblivious transfer cannot be information-theoretically secure even when combining (without further assumptions) the laws of quantum mechanics and special relativity~\cite{colbeck2007impossibility}.

\subsubsection{Now --- the free-energy-bounded model:}
We add to this list the novel \emph{free-energy-bounded model}.
Unlike the assumptions in memory-bounded models, thermodynamics does not in principle prohibit free-energy-bounded players from computing on memories of exponential size (in some security parameter), but it \emph{does} prohibit those players from \emph{erasing} a significant portion of such memories. If the players only have access to memories in \emph{initial states of maximal entropy}, as is assumed in equilibrium in thermodynamics, the erasing restriction becomes a \emph{copying} restriction (because one cannot copy without a blank memory to write the copy onto) and opens the way to a novel foundation of physics-based information-theoretic security that is different from the bounded-storage model.\footnote{In particular, the free-energy-bounded model offers fresh mechanisms, coming from reversible computing, to build information-theoretic protocols (e.g., our oblivious-transfer protocol). Another important difference is that in our protocols, the advantage of honesty in free-energy consumption is exponential in the security parameter, while in the bounded-storage model (which is not based on reversible computing but arguably more practical), it is polynomial.}

\vspace{-0.15cm}
\subsection{Reversible computing}\label{sectionthermo}
\vspace{-0.05cm}
\subsubsection{The cost of computation.}
Security in cryptography relies on a cost discrepancy between honest and malicious actors. While fundamental thermodynamical limits to the cost of computation have been well-studied (for example, see~\cite{faist2015minimal} for a quantum-informational analysis and~\cite{baumeler2019free} for an algorithmic-information-theoretical analysis), they have never before\footnote{Let us mention the (questionable) conjecture in \cite{hungerbuhler2003one} that the heat-flow equation of thermodynamics is a computational one-way function.} been considered as a means for cryptography --- we address that.

\vspace{-0.05cm}
\subsubsection{The second law of thermodynamics.}

The modern view of the second law of thermodynamics is due to \emph{Ludwig Boltzmann}, who defined \emph{the entropy of a macrostate}~--- roughly speaking, the natural logarithm of the number of microstates in the
macrostate in question~--- and stated that the entropy of a closed system does not decrease with time. The second law has constantly been subject to discourse, confusion, and dispute; its most serious challenge was ``\emph{Maxwell's demon}'' who apparently violates the law by adaptive acts, \emph{i.e.}, by a sorting procedure.
\emph{Charles Bennett}~\cite{bennett1987demons} explained that Maxwell's paradox actually disappears when the demon's internal state (its ``brain'') is taken into consideration. More specifically, \emph{the erasure} of the stored information requires free energy that is then dissipated as heat to the environment. This is \emph{Landauer's principle}~\cite{landauer61}; it did not only help to resolve the confusion around Maxwell's demon, but turned out to be an important manifestation of the second law with respect to information processing in its own right: Erasure of information~--- or, more generally, any logically irreversible computing step, has a  thermodynamic cost. \emph{Logical} irreversibility (information is lost)
implies \emph{thermodynamic} irreversibility (free energy is ``burnt'' to heat up the environment).\\

\begin{tcolorbox}[title=Landauer's principle.]
Erasing $n$ random bits requires to transform at least $n \cdot k_{\rm B} {\rm T} \ln 2 ~{\rm J}/{\rm K}$ of free energy into heat, which is dissipated into the environment.
\end{tcolorbox}

\subsubsection{Energy-neutral (thermodynamically reversible) computation.}
Landauer's principle serves as a strong motivation to ask for the possibility whether computing
can always be (made) \emph{reversible}, \emph{i.e.}, forced to not ``forget'' along the way any information about
the past (previous computation). More specifically, can every Turing-computable function
also be computed by a reversible Turing machine~(the latter was introduced in~\cite{lecerf1963machines}; see Chapter~5~of~\cite{morita2017theory} for a more modern account)? In the early 1970s, \emph{Charles Bennett}
answered this question to the affirmative; the running time is also at most
doubled, essentially~--- a very encouraging result~\cite{bennett73}: The imperative
reversibility of microphysics can, at least in principle, be carried over to macrocomputing. Bennett's
idea was that the reversible Turing machine would allocate part of its tape to maintain a history of its computation. While the latter
needs to be gotten rid of in order to have the whole be ``sustainable,'' that cannot be done by
``crude'' erasure of that history~--- all won would be lost again. It can, however, be done by \emph{un-computing}: After
copying the output, the reversible Turing machine reverts step by step the original computation, undoing its history tape in a
``controlled'' and reversible way until the output is computed back to the input. An idea similar to
Bennett's elegant trick also works for circuits: Any irreversible circuit can be transformed into
a reversible one, computing the same function, and having essentially only double depth.

All in all, this means that logical reversibility~--- which Landauer tells us to be a \emph{necessary}
condition for thermodynamic reversibility~--- can be achieved; remains the question whether it is also
a \emph{sufficient} condition for energy-neutral computation. The answer is \emph{yes}, as exemplified
by \emph{Fredkin and Toffoli}~\cite{fredkin1982conservative} and their \emph{Gedankenexperiment} of a ``ballistic computer''
which carries out its computations through elastic collisions between balls and balls, and balls and walls.

In the end, we get an optimistic picture for the future of computing: \emph{Any computable function can be computed also without
the transformation of free energy into heating of the environment}. (Clearly, a ``loan'' of free energy is necessary to start the computation, but no law of physics prevents its complete retrieval, alongside the result of the computation, when the latter concludes.)\\

\begin{tcolorbox}[title=Reversible computing.]
Any logically reversible computation can be done at zero free-energy cost by a reversible Turing machine.
\end{tcolorbox}

Reversible computing is at the core of our model.\footnote{Reversible computing is of paramount importance in the context of Moore's and Koomey's laws about the future of computation, because their continuation is threatened by physical walls and the most important one comes from thermodynamics (and not quantum mechanics). Reversible computing can in principle solve the problem completely by enabling computation without dissipation of heat. }

\subsubsection{The energy value of redundancy.}\label{szilardsection}
The converse of Landauer's principle states that all physical representations of the all-0 string have work value. More generally, all redundant (i.e., compressible in a lossless fashion) strings have work value, which is essentially their length minus their best compression~\cite{bennett1982thermodynamics}. A bound in free energy is therefore a bound on the redundancy of information; a principle we use in this work to construct cryptographic protocols.

\begin{figure}[htbp]
\begin{center}
\includegraphics[width=1\textwidth]{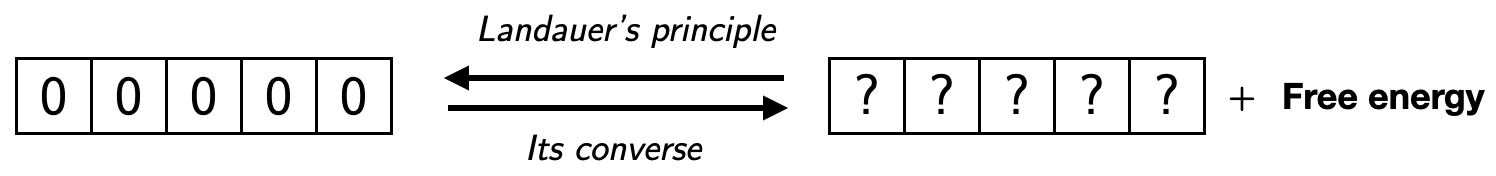}
\caption{Given the existence of thermodynamical heath baths, there is a fundamental equivalence between free energy and redundancy (\emph{i.e.}, the absence of randomness).} \label{figure1}
\end{center}
\end{figure}

\begin{tcolorbox}[title=The converse of Landauer's principle.]
It is possible to extract an amount $n \cdot k_{\rm B} T \ln 2$ of free energy from an environment by randomizing $n$ blank bits.\end{tcolorbox}

In the light of Landauer's principle and of its converse, the all-0 string can be used as a proxy for free-energy (see Fig.~\ref{figure1}). This allows us to abstract the thermodynamics completely from the model we present in Section~\ref{sectionmodel}, which is then formulated purely in terms of (logically reversible) Turing machines.

\section{Turing Machines with Polynomial Free-Energy Constraints}\label{sectionmodel}
In the following, we have this classical\footnote{The classical setting is used for all sections but Section~\ref{toquantum}, which approaches the quantum generalization.} setting in mind: Alice, Bob, and Eve have their own secure labs, where they can store and manipulate exponentially long (in some security parameter $\nu$) bit strings. Those strings start in uniformly random\footnote{This randomness is motivated by the equipartition assumption of classical thermodynamics.} states; we can think of them as the information about the specific microstate that describes the position and momentum of an exponential number of particles floating in their labs. We assume that technology is advanced enough to consider these exponentially long bit strings as static (even if the system starts in a random state, it does not get re-randomized at every time step), either because their evolution is tractable (it evolves according to the logically reversible laws of physics) or because the players can act on them quickly enough that it does not matter. The physical restriction on the honest and malicious players concerns their available free energy: For some security parameter $\nu$, malicious players are bounded exponentially (more precisely, by $2^\nu$), while honest players need only an asymptotically $\bigO(\nu)$ amount. These bounds are constraining because any computation that is not logically reversible has a free-energy cost; a malicious agent cannot for example erase a $2\cdot 2^{\nu}$-long segment of random information --- by Landauer's principle, doing so would cost a quantity of free energy exceeding their free-energy bound. We formalize this computation model in Section~\ref{computationmodel}.

Communicationwise, the players are allowed to broadcast $\bigO(\nu)$-length bit strings in the traditional sense using a public authenticated channel, or to transfer $\bigO(2^\nu)$-long bit strings through a private-but-insecure\footnote{By ``insecure,'' we mean here that it is vulnerable to Eve-in-the-middle attacks.} $\textsc{SWAP}$ channel, This channel, which swaps two bit strings at no energy cost, can also be substituted by an insecure \emph{physical} channel. Both views are informationally equivalent, and are defined in Section~\ref{reversiblecomm}.

In particular, our model differs from the bounded-storage
model --- both the players and the adversary have more power.

\subsection{Computation model}\label{computationmodel}
The fundamental laws of physics are logically reversible. We hence base
our formal notion of player (or adversary) on reversible Turing machines.\newcommand{\pfe}[0]{{\rm\textbf{PFE}}}
\newcommand{\bfe}[0]{{\rm\textbf{BFE}}}

\begin{definition}[TTM]\label{TTM}
A \emph{thermodynamical Turing machine} (TTM) is a logically reversible, deterministic, universal, prefix-free Turing machine with the following semi-infinite tapes:
\begin{enumerate}\rm
\item An input-only \textbf{instruction} tape.
\item An initially blank \textbf{computation} tape that must be returned blank when the machine halts.
\item An initially random \textbf{memory} tape.
\item An initially blank \textbf{free-energy} tape.
\end{enumerate}
\end{definition}

The \textbf{free-energy} tape of a TTM imitates a ``reservoir'' of free energy:

\begin{definition}[consumption]
 The \emph{free-energy input} $w_{\rm in}$ is quantified\footnote{More precisely, it is bounded from below.}, when the machine halts, by the distance, on the initially blank \textbf{free-energy} tape, between the extremity and the last cell with a $1$ (after this cell, the tape contains only 0s). 
\end{definition}

For example, if a machine always manages to return the \textbf{free-energy} tape as blank as it was --- it uses no free energy and computes both logically and thermodynamically reversibly; if a machine writes, and leaves, some information on the first $n$ cells of the initially blank \textbf{free-energy} tape, we say it \emph{consumes} an amount $w_{\rm in}=n$ of free-energy. ({In this work we have set $k_{\rm B} {\rm T} \ln2 \coloneqq 1$.})

Our security proofs will rely on a concept we name \textbf{proof-of-work}.
\begin{definition}[production]
We say a TTM produces a \emph{\textbf{proof-of-work}} of value $w_{\rm out}$ if it halts with a number $w_{\rm out}$ of $0$s at the beginning of its (initially random) \textbf{memory} tape.
\end{definition}

We consider agents (TTMs) with bounds, in the security parameter $\nu$, on the free-energy input.

\begin{definition}[\bfe]\label{def:pfe}
An $f(\nu)$-\bfe~agent --- an agent who is \emph{bounded in free energy by the function $f(\nu)$}, where $\nu$ is a security parameter --- is modelled by a TTM that can only consume a quantity $f(\nu)$ of free energy.
\end{definition}
In other words, every time a $f(\nu)$-\bfe~agent reaches a halting state, the non-blank portion of its \textbf{free-energy} tape ends at a distance at most $f(\nu)$ from the extremity, by definition.

In our protocols, the honest players are asymptotically $\bigO(\nu)$-\bfe, while the adversary is assumed exactly $2^{\nu}$-\bfe.
An important limitation of $f(\nu)$-\bfe~agents is given by the following theorem, to which the security of our protocols will be reduced.
\begin{theorem}\label{thm:violation}
For all $k>0$, an $f(\nu)$-\bfe~player cannot produce an $f(\nu)+k$ \textbf{proof-of-work}, except with probability $2^{-k}$.
\end{theorem}

The theorem is a consequence of the logical-reversibility characteristic imposed by the second law of thermodynamics. The proof is done in Section~\ref{app:violation}, based on Definitions~\ref{TTM}~and~\ref{def:pfe} (\emph{i.e.}, with no further references to thermodynamics).

\subsection{Communication and reversible transfer}\label{reversiblecomm}
Our cryptographic model can be formalized further by integrating \bfe~parties into a multi-round interactive protocol that uses reversible computing, however, let us focus on how Alice and Bob can exchange information. There are of two distinct resources: 
\begin{itemize}\rm
\item Standard communication for messages of length $\bigO(\nu)$.
\item Reversible transfer for longer messages, up to length $\bigO(2^\nu)$.
\end{itemize}

\subsubsection{Standard communication.}
We consider that Alice and Bob have access to a \emph{public} \emph{authenticated} communication channel in the traditional sense: Alice broadcasts a message (making, therefore, inevitably many copies of its information content) and Bob receives it. Because Alice and Bob are $\bigO(\nu )$-$\bfe$, this information-duplicating channel can only be used for messages of length $\bigO(\nu)$.

\subsubsection{Reversible transfer.}
To send states of length more than $\bigO(\nu )$, Alice and Bob have to resort to reversible computing. Reversible transfer differs from standard communication in the sense that, in order to implement the process at no free energy cost, the sender \emph{must forget} the information content of the message they send. (They could, of course, pre\"emptively make a partial copy of that information, but copying is not free and is thus limited by the free energy assumption.)
There are two different physical ways to picture such reversible transfer.

The first way is to implement, over a given distance, a reversible \textsc{SWAP}: In essence, this operation simply swaps two bit strings of equal length in a logically and thermodynamically reversible way --- Alice gets Bob's string and Bob gets Alice's string. Since we are only interested in the string that Alice (the sender) sends, Bob (the receiver) can input junk in exchange. The \textsc{SWAP} allows $\bigO(\nu )$-$\bfe$ players to transfer between themselves $\bigO(2^\nu)$ bits of information (without copying them).

The second way to implement reversible communication is to simply consider that Alice is sending the whole physical system encoding her string (\emph{e.g.}, she puts a canister of gas with entropy $2^\nu$ on a frictionless cart and pushes it toward Bob). For the cart as for the \textsc{SWAP} channel, since the information is never copied, it can be transferred from Alice to Bob at no thermodynamical cost. This is not dissimilar to how it is in practice cheaper to send hard drives directly by mail rather than to send their content through a cable.

These two pictures (the \textsc{SWAP} channel and the physical channel) are from an information point of view equivalent --- we adopt the \textsc{SWAP} channel for this work.

\section{Technical Preliminaries}\label{secproof}
We introduce some notation and introduce some of the techniques used later in the security proof of our main protocols.

\subsection{Smooth min-entropy}

Most of our formal propositions rely on the \emph{variational distance}. 

\begin{definition}
The \emph{variational distance} between two random variables $X$ and $Y$ is defined as 
\begin{equation}
\delta(X,Y)\coloneqq \frac{1}{2} \sum_{i \in \mathcal{X}\cup \mathcal{Y}} \left|p(X=i)-p(Y=i)\right|\,.
\end{equation}
It is operationally very useful because it characterizes the impossibility to distinguish between $X$ and $Y$ --- using any physical experiment whatsoever. More precisely, given either $X$ or $Y$ with probability $1/2$, the optimal probability to correctly guess which one it is is $(1+\delta(X,Y))/2$.
\end{definition}

\begin{definition}
 The \emph{conditional min-entropy} $H_\infty(X|Y)$ is defined as
\begin{equation} H_\infty (X|Y)\coloneqq -\log \sum_y P(Y=y)  \max_x  P(X = x|Y=y)\,.\end{equation}
\end{definition}
It is the optimal probability of correctly guessing $X$ given side information $Y$.

Smoothing entropies~\cite{renner2004smooth,renner2005simple} is done to ignore events that are typically unlikely. We will typically use smoothing with a parameter $\epsilon= \negl{\nu}$. We denote by $\negl{\nu}$ the functions that are \emph{negligible} in $\nu$, meaning asymptotically bounded from above by the inverse of every function that is polynomial in $\nu$.
\begin{definition}
The \emph{smooth conditional min-entropy} $H^\epsilon_\infty (X|Y)$ is defined as

\begin{equation}
H^\epsilon_\infty (X|Y) \coloneqq \max_{\omega \in \Omega \mbox{~s.t.~}P(\omega)\ge1-\epsilon} \min_y \min_x ( - \log P(X=x|Y=y,\omega)  ) \,,
\end{equation}
where $\Omega$ is the set of all events.
\end{definition}
Smooth conditional min-entropy is used mainly for privacy amplification.

\subsection{Proof of Theorem~\ref{thm:violation}}\label{app:violation}
We define and prove formally a version of Landauer's principle (Theorem~\ref{thm:violation}), which is the claim in Section~\ref{sectionmodel} that $\bfe$ players modelled as thermodynamical Turing machines cannot produce more free energy than they consume, except with exponentially vanishing probability. The theorem follows from the logical reversibility of a TTM --- the existence of a thermodynamically free logically irreversible physical process would be a violation of the second law of thermodynamics. We introduce some algorithmic-information-theory notation along the way; a more exhaustive introduction is the excellent book by Li and Vit\'anyi~\cite{li2008introduction}.

\setcounter{theorem}{0} \begin{theorem}[technical]\label{secondlaw}
Given infinite tapes $\{x,y\}$, a $f(\nu)$-\bfe~TTM $U_p(x,y)$ cannot produce a $f(\nu)+k$ \textbf{proof-of-work}, except with probability $2^{-k}$.
\end{theorem}

$\{p,x,y\}$ are, respectively, the representation of the \textbf{instruction}, \textbf{memory}, and (blank) \textbf{free-energy} tapes, at the beginning of the computation.

We start with the simpler case of assuming that all of these tapes are finite (but arbitrarily long), and then generalize our analysis to the infinite case.

\subsubsection{The finite case.}

Let $U_p(x,y)$ be a thermodynamical Turing machine as described in Definition~\ref{TTM}:  universal, prefix-free, deterministic and logically reversible. The program $p$ is taken from the read-only \textbf{instruction} tape (which can be taken long but finite); the (initially random) \textbf{memory} tape starts in $x\in_R\{0,1\}^{\len{x}}$, with $\len{x}$ taken arbitrary but finite; the \textbf{free-energy tape} starts with blank content $y=0^{\len{y}}$, where $\len{y}$ is also finite.

The logical-reversibility condition means $U_p(x,y)=U_p(x',y')$ if and only if $(x,y)=(x',y')$.

We use a counting argument. We consider the set $S$ of all couples $(x,y)$ of lengths fixed. There are $\#S=2^{\len{x}}$ of them and they are all equally probable. We then consider the subset
\begin{equation}
S(w_{\rm in},w_{\rm out}) \coloneqq \left\{ x,y {\rm ~s.t.~} U_p(x,y)=\tilde{x},\tilde{y}{\rm ~with~} \begin{cases}
\tilde{x}=0^{w_{\rm out}}\,||\,*\\
\tilde{y}=*\,||\,0^{\len{y}-w_{\rm in}}
\end{cases}\right\}\,,
\end{equation}
where $*$ is an arbitrary padding string of appropriate length, and $\,||\,$ denotes a concatenation.
Intuitively, $w_{\rm in}$ bounds the free-energy input and is the minimum number of bits that get randomized on the initially blank \textbf{free-energy} tape $y$; $w_{\rm out}$ bounds the free-energy output and is the maximum number of erased bits on the initially random \textbf{memory} tape $x$. (Those erased bits constitute the \textbf{proof-of-work}.)

\begin{lemma}
\begin{equation}
\# S(w_{\rm in},w_{\rm out})\le 2^{\len{x}-{w_{\rm out}}+{w_{\rm in}}}\,.\end{equation}
\begin{proof}
Because of logical reversibility, the input-couples $(x,y)\in S$ are at most\footnote{``At most'' because not all programs halt and some output-couples might not be in the image of $U_p$.} as numerous as the output-couples 
$
(\tilde{x},\tilde{y}) \mbox{~s.t.~} \begin{cases}
\tilde{x}=0^{w_{\rm out}}\,||\,*\\
\tilde{y}=*\,||\,0^{\len{y}-w_{\rm in}}
\end{cases}
$
\hspace{-12pt}. We count the maximum number of such output-couples by summing the lengths of all ``$*$ positions''; there are at most $2^{(\len{x}-w_{\rm out})+w_{\rm in}}$ of them.\qedhere
\end{proof}
\end{lemma}

The probability of drawing at random such a couple $(x,y)$ is therefore
\begin{equation}P(x,y\in S(w_{\rm in},w_{\rm out}))\le \# S(w_{\rm in},w_{\rm out}) / \#S=2^{w_{\rm in}-w_{\rm out}}\,.\end{equation}

\begin{proposition}\label{proposition6}
Given finite $\len{x}$ and $\len{y}$, a $f(\nu)$-\bfe~TTM $U_p(x,y)$ (therefore with free-energy input $w_{\rm in}=f(\nu)$) is limited in its production of free energy $w_{\rm out}$ by
\begin{equation}
\forall k>0,~ P\left( w_{\rm out} > w_{\rm in}+k \right) \le 2^{-k} \,.\end{equation}
\end{proposition}

\subsubsection{The infinite case.}
We now reduce the infinite case to the finite case that we just analyzed.

We take again a TTM. Let us consider $x\in_R\{0,1\}^\infty$, where each bit is perfectly random. Let us also set $y=0^\infty$. Since $p$ is fixed, it is enough to again consider it finite. The prefix-free condition implies that the behaviour of $U_p(x,y)$ is well defined even on infinite tapes because its programs\footnote{``Program'' is taken here in the general sense and includes arguments $p$ and $x$.} are self-delimited.

\newcommand{\prefix}[1]{\textit{\rm effective}(#1)}

\begin{definition}
Let \begin{equation}\displaystyle \Omega_{U_p}\coloneqq \sum_{\prefix{x} \text{~s.t.~} U_p(x,y) \text{~halts}} 2^{- \len{\prefix{x}}} \end{equation} be the \emph{halting probability} of $U_p$ (\emph{i.e}, Chaitin's constant~\cite{chaitin1975theory}), where the sum is over all self-delimited programs $\prefix{x}\in \{0,1\}^*$ \footnote{We assume $p$ to be fixed; by ``program'' we mean the random input $x$.}.
\end{definition}

We also define its partial sum.
\begin{definition}
\begin{equation}
\displaystyle{\Omega_{U_p}}(n)\coloneqq \sum_{\prefix{x} \text{~s.t.~} U_p(x,y) \text{~halts~and~} \len{\prefix{x}}\le n } 2^{- \len{\prefix{x}}}.\end{equation} 
\end{definition}

Note first that since ${\Omega_{U_p}}(n)$ is a monotonically increasing function that converges to ${\Omega_{U_p}}$, it holds that
\begin{equation}
\forall \epsilon > 0, \exists N' {\rm ~s.t.~} \Omega_{U_p}- \Omega_{U_p}(N') < \epsilon\,. \label{mono}\end{equation}

\begin{definition}
Let ${\rm BB}_{U_p}(n)$ be the \emph{time-busy-beaver function}, which returns the maximum running time that a halting program $\prefix{x}$ of length $\le n$ can take before halting.\end{definition}
Observe that it implies that, for all halting programs of length $\le n$, the infinite part of each tape that comes after the $({\rm BB}_{U_p}(n))^{\rm th}$ bit is never read or modified by the TTM (moving there is by definition too long).

\begin{proposition}\label{ttm2}
A TTM with infinite tapes $(x,y)$ behaves with arbitrarily high probability exactly as if these infinite tapes were (extremely long but) finite:\\ $\forall \epsilon>0$, $\exists N$ such that \begin{equation}
P\left( {U_p}(x,y)=\left( U_p(x_{[\le N]},y_{[\le N]})~\,||\,~(x_{[> N]},y_{[> N]}) \right)  \right) \ge 1- \epsilon \,,
\end{equation}
where the subset notation is used to split $x=x_{[\le N]}\,||\,x_{[> N]}$ and $y=y_{[\le N]}\,||\,y_{[> N]}$.
\begin{proof}
Taking Eq.~\ref{mono} with $N \coloneqq{\rm BB}_{U_p}(N')$, with the consideration about busy beaver above (any machine that halts affects only a finite amount of tape).\qedhere
\end{proof}
\end{proposition}

Finally, Theorem~\ref{secondlaw} is obtained by combining Proposition~\ref{proposition6} and Proposition~\ref{ttm2}, with $\epsilon \rightarrow 0$:
\begin{equation}
\forall k>0,\, P\left(w_{\rm out} > f(\nu)+k\right) \le 2^{-k} \,,
\end{equation}
where $w_{\rm out}$ is the value of the \textbf{proof-of-work}.

\subsection{The exhaustive and sampled memory games}\label{memorygames}

We detail here in a game format a reduction that we later use in our security proofs. Our memory games involve an adversary against a verifier. The adversary sends, using a reversible channel \textsc{SWAP}, an exponentially long string to the verifier, but is also asked to try to keep a copy of it; the verifier then interrogates the adversary about either all of that string (in the \emph{exhaustive} variant), or about a random linear-size subset of it (in the \emph{sampled} variant); we show that the adversary has limited advantage in guessing as compared to a trivial strategy, unless they made an accurate copy of the whole string of exponential length --- a process that requires, in light of Landauer's principle, an exponential amount of either luck or free energy. We formalize this intuition, starting with the non-sampled version of the game.

\begin{definition}
The \emph{exhaustive $k\cdot2^{\nu} \choose k\cdot2^{\nu}$ memory game} is defined as follows for security parameters~$\nu$ and $k$:
\begin{enumerate}
\rm
\item The adversary isolates (by taking it from the environment of their lab for example) a system $X\in \mathcal{X}= \{0,1\}^{k\cdot2^{\nu}}$. All the rest of their available information is modelled as $E$.
\item The adversary (modelled as a TTM) makes some computation on the systems $X,E$.
\item Through a noiseless reversible channel (\emph{e.g.}, \textsc{SWAP}), the adversary sends $X$ to the verifier.
\item The verifier provides the adversary a blank tape of length $k\cdot2^{\nu}$, and asks the adversary to correctly print on it all of $X$.
\end{enumerate}
\end{definition}

\begin{proposition}\label{oneprop}
For any $2^\nu$-\bfe~adversary, the advantage at the exhaustive $k\cdot2^{\nu} \choose k\cdot2^{\nu}$ memory game, compared to a trivial coin-flip strategy, is bounded by
\begin{equation}
H_\infty(X|E)\ge (k-1)2^\nu\,.
\end{equation}

\begin{proof}
We reduce a violation of Theorem~1 (\emph{i.e.}, Landauer's principle) to a large advantage at the exhaustive $k\cdot2^{\nu} \choose k\cdot2^{\nu}$ memory game.
During the game, instead of sending $X$ to the verifier, the adversary deviates and XORs onto $X$ their best guess for $X$ given side information $E$. If the adversary guesses correctly, it turns $X$ into an all-$0$ string. This \textbf{proof-of-work} of length $k\cdot2^{\nu}$ violates Theorem~1 if it is created with probability higher than $2^{-(k-1)2^{\nu}}$; therefore, it does not.\qedhere
\end{proof}
\end{proposition}

The constraint also holds if the adversary is quizzed only on a random subset of positions.
\begin{definition}
The \emph{sampled $k\cdot2^{\nu} \choose t$ memory game} is defined as follows for free-energy bound $2^\nu$, security parameter $k$, and sample size $t$:
\begin{enumerate}
\rm
\item The adversary isolates (by taking it from the environment of their lab for example) a system $X\in \mathcal{X}= \{0,1\}^{k\cdot2^{\nu}}$. All the rest of their available information is modelled as $E$.
\item The adversary (modelled as a TTM) makes some computation on the systems $X,E$.
\item Through a noiseless reversible channel (\emph{e.g.}, \textsc{SWAP}), the adversary sends $X$ to the verifier.
\item The verifier chooses at random $t$ \textit{sample} positions $\subset\mathcal{X}$ and sends a description of these positions to the adversary, who must correctly guess $X_{[\textit{sample}]}$.
\end{enumerate}
\end{definition}

\begin{theorem}\label{twothm}
For any $2^\nu$-\bfe~adversary, the advantage at the sampled $k\cdot2^{\nu} \choose t$ memory game, compared to a trivial coin-flip strategy, is bounded, for all $\delta>0$, by
\begin{equation}
H_\infty^{\negl{t}}(X_{[\textit{sample}]}|E)\ge \frac{t\cdot (k-1)}{k} -t\cdot \delta \,.
\end{equation}

\begin{proof}
Lemma~6.2~in~\cite{vadhan2004constructing} states that, under random sampling, the min-entropy per bit is with high probability approximately conserved. In our case, this implies that, for all $\delta>0$,
\begin{equation}
H^{2^{-\Omega(t\delta^2 \log^2{\delta})}+2^{-\Omega(k 2^\nu \delta )}}_\infty(X_{[\textit{sample}]}|E) \ge \frac{t}{ k\cdot2^{\nu}} H_\infty(X|E)-t\cdot \delta\,,
\end{equation}
given which Theorem~\ref{twothm} follows from Proposition~\ref{oneprop}.\qedhere \end{proof}
\end{theorem}

\subsection{Universal hashing}\label{universalhash}
Universal hashing is useful for both privacy amplification and authentication.

\begin{definition}[2-universal hashing~\cite{carter1979universal,wegman1981new}]
Let $\mathcal{H}$ be a set of hash functions from $\{0,1\}^n \rightarrow \{0,1\}^m$. $\mathcal{H}$ is \emph{2-universal} if, given any distinct elements $x_1,x_2 \in\{0,1\}^n $ and any (not necessarily distinct) elements $y_1,y_2\in\{0,1\}^m$, then 

\begin{equation}
\# \{h\in \mathcal{H}| y_1{=}h(x_1)\land y_2{=}h(x_2)\}=\#\mathcal{H}/ 2^{2m} \,.
\end{equation}
\end{definition}

\begin{lemma}[Leftover hash lemma~\cite{bennett1988privacy,impagliazzo1989pseudo,haastad1993construction,bennett1995generalized}]\label{leftover}
Let $h: \mathcal{S} \otimes \mathcal{X}  \rightarrow \{0,1\}^m$ be a 2-universal hash function. If $H_\infty(X) \ge m + 2 \epsilon $, then
\begin{equation}
\delta \Big( (h(S,X),S) , U\otimes S \Big) \le  {2^{-\epsilon}}\,.
\end{equation}
$S$ is a short uniformly random seed and $X$ is the variable whose randomness is to be amplified. $U$ is the uniform distribution of appropriate dimension. The symbol $\otimes$ is used to represent the joint probability of independent distributions.
\end{lemma}

\section{Secret-Key Establishment}\label{secske}
Secret-key establishment (SKE) is a fundamental primitive for two-way secure communication because it allows for a perfectly secure one-time-pad encryption between Alice and Bob about which Eve knows nothing (otherwise the protocol aborts).
\subsection{Definitions (SKE)}

\begin{definition}
A secret-key-establishment scheme is \emph{sound} if, at the end the protocol, Alice and Bob possess the same key with overwhelming probability in the security parameter $\eta$:
\begin{equation}
P(K_A \neq K_B)\le {\negl{\eta}}\,.
\end{equation}
\end{definition}

\begin{definition}
A secret-key-establishment scheme is information-theoretically \emph{secure} (\emph{i.e.}, almost perfectly secret) if the key $K_B$ is uniformly random even given all of the adversary's side information $E$, except with probability at most negligible in the security parameter $\nu$:
\begin{equation}
\delta \Big( (K_B,E), U \otimes E \Big) \le {\negl{\nu}}\,.
\end{equation}

\end{definition}
In what follows, the variables $(A,B)\in (\mathcal{A},\mathcal{B})$ are strings from registers of length roughly $\bigO(\nu \log \nu)$, while
$(X,Y)\in (\mathcal{X},\mathcal{Y})$ denote strings from registers of length $\bigO(2^\nu)$.

\subsection{Protocol (SKE)}\label{protocolske}
\begin{theorem}
The following secret-key-establishment protocol is information-{\linebreak}theoretically sound and secure against any eavesdropper whose free energy is bounded by $2^\nu$. Alice and Bob need a quantity of free energy that is asymptotically $\bigO(\nu)$. \end{theorem}
Soundness is analyzed in Section~\ref{SKE sound}, and security in Section~\ref{skesecure}.

\begin{tcolorbox}[pad at break*=1mm,
  colback=gray!10!white,colframe=gray!75!black,title={Secret-key-establishment protocol:}]

\begin{enumerate}
	\item Alice starts \footnote{
The main parameters are\\
{- $\nu$, from the $2^\nu$ bound in free energy of Eve;\\
- $k$, which determines the tolerated error rate between Alice and Bob;\\
- $t$, the number of test bits to estimate the above error rate;\\
-~$s$, the length of the raw key (before processing).
}} with $X\in \mathcal{X}=\{0,1\}^{k\cdot 2^{\nu}}$ in a uniformly random state (extracted from the equidistributed environment of her lab). She draws uniformly at random a subset $\subset \{1,\dots,{k\cdot 2^{\nu}}\}$ of $s+t$ positions ${\it rawkey} $ and copies $({\it rawkey},X_{[{\it rawkey}]})\rightarrow A$ to her memory.
	\item Alice sends ${X}\rightarrow {Y} $ to Bob using a reversible channel ({\it e.g.}, a \textsc{SWAP} channel); it is possibly intercepted by Eve.\label{step2}
	\item Bob announces the receipt to Alice on an authenticated public channel. In case of no  receipt, they abort.\label{step3}
	\item Alice publishes the subset positions ${\it rawkey}$ on the (noiseless) authenticated public channel so that Bob can select $Y_{[{\it rawkey}]}\rightarrow B$. Alice and Bob draw a ${\it test}$ sub-subset of $t$ bits that they sacrifice to estimate the error rate $p_{\rm error}$ between $A$ and $B$.
	\item If the estimated $p_{\rm error}$ is too large, they abort. Otherwise, Alice and Bob apply information reconciliation (detailed in Section~\ref{SKE sound}) on the remaining $s$ bits $A_{[{\overline{\it test}}]}$ and $B_{[ {\overline{\it test}}]}$. 
	\item Alice and Bob apply privacy amplification (detailed in Section~\ref{skesecure}) and obtain a shared secret key of length $\approx ((k-1)/k- h_b(p_{\rm error}))\cdot s$.
\end{enumerate}
\end{tcolorbox}

$h_b(p)\coloneqq -p\log_2 p - (1-p) \log_2 (1-p)$ is the \emph{binary entropy}.

Note that for any fixed $p_{\rm error}$ (as long as it is not trivially $1/2$), Alice and Bob can choose a security parameter $k$ for which the protocol will be secure for that value of $p_{\rm error}$. That is unlike, for example, the BB84 quantum-key-distribution protocol, which only tolerates error rates less than $1/4$ (any more and Eve can intercept the whole quantum state).

\subsubsection{The intuition.}
Because she is $2^\nu$-bounded in free energy, Eve cannot copy to her memory the whole $k\cdot 2^{\nu}$-long string $Y$ that she sends to Bob, on which Bob will later base the raw key. Alice circumvents this limitation by already knowing the raw-key positions at the moment she sends $X$ ($X$ becomes, after Eve's potential tampering, $Y$) and thus need not store more than an asymptotically $\bigO\left(\nu\right)$-long segment of the $k\cdot 2^{\nu}$-long string. As in quantum key distribution, Eve can force the protocol to abort.

\subsection{Soundness analysis (SKE)}\label{SKE sound}
\subsubsection{Parameter estimation.}
We first estimate (using upper bounds) between Alice and Bob the global error rate $p_{\rm error}$ and the non-tested {\it rawkey} error rate $p_{\rm error}^{\overline{\it test}}$. The former quantity is important for the privacy amplification analyzed in Section~\ref{skesecure}, while the second is needed to analyze information reconciliation.

\begin{proposition}
Alice and Bob can accurately estimate the error rate $p_{\rm error} $ by sampling on the $t$ ${\it test}$ positions the error rate $p_{\rm error}^{\it test}$:
\begin{equation}
P\left( p_{\rm error}  \le p_{\rm error}^{\it test} +  \epsilon \right) \ge 1-{ e^{-2 \epsilon^2 t}}\,.\label{reshuffling}
\end{equation}
\begin{proof}
$p_{\rm error}^{\it test}$ is computed from the Hamming weight $\omega(\overline{{A_{[{\it test}]}\oplus B_{[{\it test}]}}})=t(1-p_{\rm error}^{\it test} )$. Chernoff's inequality bounds $p_{\rm error}$.\qedhere
\end{proof}
\end{proposition}

\begin{proposition}
Alice and Bob can accurately estimate $p_{\rm error}^{\overline{\it test}} $ from $p_{\rm error}^{{\it test}} $:
\begin{equation}
P\left( p_{\rm error}^{\overline{\it test}}  \le p_{\rm error}^{{\it test}} + \frac{s\cdot \epsilon}{s+t}\right) \ge 1-{ e^{-2 \epsilon^2 t}}\,.
\end{equation}
\begin{proof}
We insert $p_{\rm error}= (s\cdot p_{\rm error}^{\overline{\it test}} + t\cdot p_{\rm error}^{{\it test}})/(s+t)$ in Eq.~\ref{reshuffling} and isolate $p_{\rm error}^{\overline{\it test}} $.\qedhere
\end{proof}
\end{proposition}

\subsubsection{Information reconciliation (error correction).}\label{correction}

Once they have a good estimate of $p_{\rm error}^{\overline{{\it test}}}$, Alice and Bob achieve information reconciliation by applying error correction on that unused subset $\overline{{\it test}}$ of $s$ bits.

Note that it is important that the established key be based on Bob's string, rather than on Alice's, because the reasoning (see the security analysis in Section~\ref{skesecure}) using the sampled memory game only directly bounds from above the mutual information between Bob and Eve, not the one between Alice and Eve.
\begin{proposition}\label{errorprop}
For any non-trivial constant $p_{\rm error}^{\overline{{\it test}}}\neq 1/2$, Alice and Bob can transform the samples $A_{[{\overline{\it test}}]},B_{[ {\overline{\it test}}]}$ into the (non-necessarily secret) keys $K'_A,K'_B$ for which
\begin{equation}
P\left( K'_A = K'_B \right) \ge 1- {\negl{\eta}}\,.
\end{equation}
They can do so with $w\approx  h_b(p_{\rm error}^{\overline{{\it test}}})\cdot s$ (the exact value is given below) bits of authenticated public communication.
\end{proposition}

We present one standard construction to correct an arbitrary error rate on the~$s$ bits of \textit{rawkey} that were not used during the parameter-estimation phase.

\paragraph{Asymptotically optimal protocol for information reconciliation~\cite{brassard1993secret}:}
~

Let $w\coloneqq  \ceil{ s\cdot  h_b(p_{\rm error}^{\overline{{\it test}}} + \delta')+\eta }$;
\begin{enumerate}
\item Bob picks at random a hash function $h:\{0,1\}^s\rightarrow \{0,1\}^w$ from a 2-universal family $\mathcal{H}$ and computes $h(B_{[\overline{{\it test}]}})$.
\item Bob communicates $h$ and $h(B_{[\overline{{\it test}]}})$ to Alice, using the authenticated public channel.
\item Alice computes $\tilde{A}_{[\overline{{\it test}]}} \coloneqq \displaystyle \argmin_{x \in \{0,1\}^{\len{s}}} \left(\omega(x,A_{[\overline{{\it test}]}})| h(x){=}h(B_{[\overline{{\it test}]}})\right)$.
\end{enumerate}

Here, $\omega(\cdot,\cdot)$ is the Hamming distance;
$\delta'$ determines efficiency and $\eta$ is the security parameter.

\begin{proof}
We first count, in the uniform distribution, the smooth number of strings with length~$s$ that contains approximately $p_{\rm error}^{\overline{\it test}}$: Let $M\coloneqq\{x \in \{0,1\}^s\,|\, p_{\rm error}^{\overline{{\it test}}} - \delta' \le p_{\rm error}^{\overline{{\it test}}}(x)\le p_{\rm error}^{\overline{{\it test}}} + \delta'\}$; from the asymptotic equipartition property, we have $\forall \delta'>0$,
\begin{equation}
P\left( \#M\le 2^{s\cdot h_b(p_{\rm error}^{\overline{{\it test}}} + \delta')} \right) \ge 1-{2^{-{ \Theta(\eta)}}}\,.
\end{equation}

Because $\mathcal{H}$ is 2-universal, the probability of obtaining a correct hash from a non-correct candidate in $M$ is bounded by $2^{-w}$. By the union bound, the protocol is therefore sound except with probability at most $2^{-w}\cdot \#M$, which is $\negl{\eta}$.\qedhere
\end{proof}

While the above ideal information reconciliation protocol is optimal, it offers no (known) efficient way (in the computational complexity sense) for Alice to decode Bob's codeword. While we are in this work only concerned with thermodynamic (rather than computational) efficiency, we refer to \cite{brassard1993secret}, or to the theory of Shannon-optimal efficient algebraic codes, such as convoluted codes, for asymptotically ideal information-reconciliation protocols that are also computationally efficient.

\subsection{Security analysis (SKE)}\label{skesecure}
If the protocol does not abort, Eve has negligible information about the key $K_B$ at the end.
This security resides on the fact that even if Eve intercepts $X$ (which was sent from Alice to Bob) and replaces it with $Y$, she cannot keep roughly more than a fraction $1/k$ of the information about~$Y$. Thus, since the key is based on $Y$, Eve has limited knowledge about it.

Formally, this can be analyzed with the sampled $k\cdot 2^{\nu} \choose s$ memory game in Section~\ref{memorygames}. Theorem~\ref{twothm} thereat guarantees a good starting point --- Eve (who is $2^{\nu}$-$\bfe$) must have limited information about Bob's raw key of length $s$:
\begin{equation}
\forall \delta >0, \,H^{\negl{\nu}+\negl{s}}_\infty(Y_{[\overline{{\it test}}]}|E,{\it rawkey},\overline{{\it test}})=s \cdot \frac{k-1}{k}-s\cdot \delta\,.\end{equation}
The next step is to go from \emph{low} information to \emph{essentially no} information.

\subsubsection{Privacy amplification.}\label{privacy}
Privacy amplification turns a long string about which the adversary has potentially some knowledge into a shorter one about which the adversary has essentially none.

In secret-key establishment, Eve's partial information can come from eavesdropping (and as shown, this quantity is roughly a fraction $1/k$) or from the public information leaked by the information reconciliation protocol, which is easily characterized.

Privacy amplification can be realized in an information-theoretically secure manner with 2-universal hashing (see Section~\ref{universalhash}).

\begin{proposition}
After privacy amplification, $K_B$ is approximately of length $\approx ((k-1)/k- h_b(p_{\rm error}))\cdot s$, and Eve has essentially no knowledge about it.
\begin{proof}
Let $w$ quantify the number of bits about $B_{[\overline{{\it test}]}}$ exchanged publicly during the information-reconciliation (IR) protocol.
We note that $H_\infty(K_B|E^{\rm pre IR})\le H_\infty(K_B|E^{\rm post IR})-w$,
hence
\begin{equation}
\forall \delta>0, \,H^{\negl{\nu}+\negl{s}}_\infty(K_B|E^{\rm post IR})=s \cdot \frac{k-1}{k}-s\cdot \delta-w \,.
\end{equation}
Therefore, taking $m\coloneqq s \cdot \frac{k-1}{k}-s\cdot \delta-w-\epsilon $ guarantees after hashing ($\epsilon$ is the security parameter for the Leftover hash lemma; see Section~\ref{universalhash}) information-theoretic security on those remaining $m$ bits.\qedhere
\end{proof}
\end{proposition}

Note that for any fixed $p_{\rm error}$, the parameters $s$ and $k$ can be selected as to make $m$ a positive quantity when the protocol does not abort (as a result of too many errors). Also note that the parameters $\nu$ and $s$ must not be too small.

\newpage

\section{1-out-of-2 Oblivious Transfer}\label{secot}
Oblivious transfer (OT) is a cryptographic primitive that is universal for two-party computation~\cite{rabin1981exchange,kilian1988founding}. It comes in many flavours, but they are all equivalent~\cite{crepeau1987equivalence}. We concern ourselves with 1-out-of-2 OT (or 1--2 OT). Informally: Alice sends two envelopes to Bob; Bob can open one to read the message in it, but he cannot open both; Alice cannot know which message Bob read.

\subsection{Definitions (OT)}
\begin{definition}
A 1--2 OT protocol is perfectly \emph{sound} if, when Alice and Bob are honest, the message $B(i)$ received by Bob is with certainty the message $m_i$ sent by Alice, for his choice of $i\in\{0,1\}$:
\begin{equation}
P\left(B (i)=m_i \right)=1\,.
\end{equation}
\end{definition}

\begin{definition}
A 1--2 OT protocol is information-theoretically \emph{secure-for-Alice} if Bob cannot learn something non-negligible about both of Alice's messages simultaneously: For any $2^{\nu}$-$\bfe$ Bob,\begin{equation}
\exists j \mbox{~s.t.~}\delta \Big( (m_{j},E_B), (U \otimes E_B)\Big) \le \negl{\eta}   \,.
\end{equation}
\end{definition}
$E_B$ denotes all of (a potentially malicious) Bob's side information. And similarly for $E_A$ in regards to Alice.

\begin{definition}
A 1--2 OT protocol is information-theoretically \emph{secure-for-Bob} if Alice cannot learn anything non-negligible about Bob's choice $i$: For any $2^{\nu}$-$\bfe$ Alice,
\begin{equation}
\delta\Big( (i,E_A), U \otimes E_A \Big) \le {\negl{\eta}}\,.
\end{equation}
\end{definition}
An OT protocol is information-theoretically secure when it is information-{\linebreak}theoretically secure for \emph{both} Alice and Bob.

\subsection{Protocol (OT)}\label{protocolot}

\begin{theorem}
The following 1--2 OT protocol is perfectly sound and information-theoretically secure against $2^\nu$-$\bfe$ adversaries. The free-energy requirement of the honest players is asymptotically $\bigO(\nu)$. \end{theorem}
The perfect soundness is straightforward. Security is analyzed in Section~\ref{OT security}.

\begin{tcolorbox}[pad at break*=1mm,
  colback=gray!10!white,colframe=gray!75!black,title={1--2 oblivious-transfer protocol:}]

(The variable $\eta$ is a security parameter.)

\begin{enumerate}
\item Alice chooses messages $m_0$ and $m_1$ of length $n$.
\item Alice starts with the exponentially long bit strings $X^{(0)},X^{(1)} \in \mathcal{X}=\{0,1\}^{4\cdot 2^{\nu}}$ in uniformly random states. She picks a random subset $\subset \{1,\dots,4 \cdot 2^{\nu}\}$ of $n+\eta$ positions ${\it raw}$ and stores $({\it raw},X^{(0)}_{[\it raw]}, X^{(1)}_{[\it raw]})$ in her memory.
\item Alice sends $(X^{(0)},X^{(1)})$ to Bob using the reversible channel \textsc{SWAP}.
\item Bob chooses $i\in \{0,1\}$ and computes reversibly $(X^{(0)},X^{(1)}) \rightarrow (X^{(i)},X^{(0\oplus1)})$, where we define $X^{(0\oplus 1)}\coloneqq X^{(0)}\oplus X^{(1)}$. Then, Bob keeps $X^{(i)}$ and sends back $X^{(0\oplus1)}$ reversibly to Alice using \textsc{SWAP}.
\item Alice receives $\tilde{X}^{(0\oplus1)}$ and checks whether $\tilde{X}^{(0\oplus1)}_{[\it raw]}{=}X^{(0\oplus1)}_{[\it raw]}$. If they differ, Alice aborts.\label{perfect match step}
\item Alice chooses at random a 2-universal hash function $h: \{0,1\}^{n+\eta} \rightarrow \{0,1\}^{n}$ and communicates $h,{\it raw},m_0 \oplus h(X^{(0)}_{[\it raw]}),m_1 \oplus h(X^{(1)}_{[\it raw]})$ to Bob.
\item Bob computes the hash $h(X^{(i)}_{[\it raw]})$ and recovers $m_i$. 
\end{enumerate}
\end{tcolorbox}

\subsubsection{The intuition.}
In addition to the previously exploited \emph{impossibility to copy} exponential quantities of information without using corresponding quantities of free energy or violating Landauer's principle, the oblivious-transfer protocol makes use of another key feature of \emph{reversible computing}: As long as Bob is in possession of $X^{(0\oplus 1)}\coloneqq X^{(0)}\oplus X^{(1)}$, the maximally random variables $X^{(0)}$ and $X^{(1)}$ have conditionally exactly the \emph{same} information content; but once $X^{(0\oplus 1)}$ is returned to Alice, $X^{(0)}$ and $X^{(1)}$ revert to being \emph{uncorrelated}. In other words, although sending $X^{(0\oplus 1)}$ back to Alice forces Bob to \emph{forget} information about the couple $X^{(0)},X^{(1)}$ (enabling 1-out-of-2 transfer), it does not uniquely specify \emph{which} information he forgot (Alice remains oblivious).

\subsection{Security analysis (OT)}\label{OT security}
\subsubsection{Security for Bob.}
From Alice's point of view, Bob's behaviour (\emph{i.e.}, sending ${X}^{(\rm 0\oplus 1)}$ back to Alice) is identical whether he chooses message $i{=}0$ or message $i{=}1$; the scheme is therefore perfectly secure for Bob.
\subsubsection{Security for Alice.}
We prove that a malicious Bob cannot learn anything non-negligible about a second message as soon as he learns something non-negligible about a first message.
\begin{proof}

We pose without a loss of generality that $\omega$ is the event corresponding to ``Bob learns something non-negligible about $m_0$.'' Because he is $2^\nu$-bounded in free energy, a malicious Bob's success at the sampled $4 \cdot 2^{\nu} \choose n+\eta $ memory game (on state $\tilde{X}^{(\rm 0\oplus 1)}$ and sample {\it raw}) is bounded by Theorem~\ref{twothm}:
\begin{equation}
\forall \delta>0,\, H_\infty^{\negl{\nu}+\negl{\eta}}(\tilde{X}^{(0\oplus1)}_{[\it raw]}|E_B,\omega)\ge (n+\eta)/2 - (n+\eta)\cdot \delta\,.\label{contra1}
\end{equation}
By subadditivity, we have
\begin{align}
&H_\infty^{\negl{\nu}+\negl{\eta}}(\tilde{X}^{(0\oplus1)}_{[\it raw]}|E_B,\omega) \\&\le H_\infty^{\negl{\nu}+\negl{\eta}}({X}^{(0)}_{[\it raw]},{X}^{(1)}_{[\it raw]}|E_B,\omega)\\ &\le H_\infty^{\negl{\nu}+\negl{\eta}}({X}^{(0)}_{[\it raw]}|E_B,\omega)+H_\infty^{\negl{\nu}+\negl{\eta}}({X}^{(1)}_{[\it raw]}|E_B,\omega) \,.
\end{align}
We apply the Leftover hash lemma (Lemma~\ref{leftover}) with $\epsilon \coloneqq \eta/12-3n/8$. The two privacy-amplification steps succeed (except by the union bound with probability $\negl{\nu}+\negl{\eta}$) if, respectively,
\begin{align}
&H_\infty^{\negl{\nu}+\negl{\eta}}({X}^{(0)}_{[\it raw]}|E_B,\omega)  \ge n/4+\eta/6\,,\\
&H_\infty^{\negl{\nu}+\negl{\eta}}({X}^{(1)}_{[\it raw]}|E_B,\omega)  \ge n/4+\eta/6 \,.
\end{align}
We assume by contradiction that they are both unsuccessful with non-negligible probability. It implies
\begin{equation}
H_\infty^{\negl{\nu}+\negl{\eta}}(\tilde{X}^{(0\oplus1)}_{[\it raw]}|E_B,\omega) < n/2 + \eta/3\,,
\end{equation}
which contradicts Eq.~\ref{contra1} for small $\delta \le \eta/(6(n+\eta))$.\qedhere
\end{proof}

\section{From classical adversaries to quantum adversaries}\label{toquantum}

Up to here, the notion of information that has been used --- in the protocols for secret-key establishment and oblivious transfer, as well as in their analyses --- is purely \emph{classical}. But as scrutinised by thorough experiments~(notably, the extensive serie of Bell experiments~\cite{freedman1972experimental,aspect1982experimental,hensen2015loophole,giustina2015significant,shalm2015strong}), nature is \emph{quantum-physical}.
The aim of this section is to bring our work one step closer to the quantum realm. Namely, we investigate whether our (classical\footnote{All classical operations can be viewed as quantum operations restricted to diagonal density matrices.}) protocols are secure against quantum adversaries. We find that our SKE protocol (Section~\ref{protocolske}) is secure against a quantum Eve \emph{as it is}. On the other hand, to retain security against a malicious quantum Alice, our OT protocol (Section~\ref{protocolot}) has to be slightly updated --- the patched protocol presented below in Section~\ref{patchedprotocol} is quantum-safe but remains classical for honest players. Our work's conclusion, therefore, fully extends to the \emph{quantum} world of Maxwell demons (given arbitrarily large but random environments): It is~--- on paper ---~information-theoretically cryptographically friendly.

\subsection{The setting made quantum}
Our model described in Section~\ref{sectionmodel} is based on Alice, Bob, and Eve being classical computers with thermodynamical restrictions (we call them Thermodynamical Turing Machines) interacting through classical channels (a standard authenticated channel and a \textsc{SWAP} channel).

In a quantum setting, Alice, Bob, and Eve are upgraded to universal quantum computers~\cite{deutsch1985quantum} and their communication channels can carry states in quantum superposition. A quantum computer cannot compute more than a classical computer could (given exponential computational time, a classical computer can simulate a quantum computer). Quantum computing cannot either be used to evade Landauer's principle~\cite{faist2015minimal}. As such, once all elements are properly defined, a quantum version of our Theorem~\ref{thm:violation} holds.

\begin{proposition}[Thm.~\ref{thm:violation} in the quantum realm (sketch)]\label{quantumsecondlaw}
For all $k>0$, a player modelled by a quantum computer with a bound $f(\nu)$ in free energy cannot erase more than $f(\nu)+k$ initially completely mixed qubits, except with probability $2^{-k}$.
\end{proposition}

The ability to send and receive quantum states does enable new possibilities for both honest and malicious agents --- we investigate next how this affects the security of our previous SKE and OT protocols.

\subsection{The quantum exhaustive and sampled memory games}
We extend the proof method developed in Section~\ref{memorygames} to the quantum world.

First, the bound on the success of an adversary at the exhaustive $k\cdot2^{\nu} \choose k\cdot2^{\nu}$ memory game (Proposition~\ref{oneprop}) is unaffected by the transition from classical to quantum information.

\begin{proposition}[Prop.~\ref{oneprop} with quantum side-information]\label{onepropQ}
For any quantum adversary with a bound $2^\nu$ in free energy, the advantage at the exhaustive $k\cdot2^{\nu} \choose k\cdot2^{\nu}$ memory game, compared to a trivial coin-flip strategy, is bounded by
\begin{equation}
H_\infty(X|E)\ge (k-1)2^\nu\,.
\end{equation}
\begin{proof}
$X$ is here still classical, but $E$ represents side information that is possibly quantum. Since the operational meaning of conditional min-entropy is the same whether the side information is quantum or not~\cite{konig2009operational}, the argument presented in Section~\ref{memorygames} is unchanged.\qedhere\end{proof}
\end{proposition}

The next step is to sample from $X$ (Theorem~\ref{twothm}).

\begin{proposition}[Thm.~\ref{twothm} with quantum side-information]\label{twothmQ}
For any quantum adversary with a bound $2^\nu$ in free energy, the advantage at the sampled $k\cdot2^{\nu} \choose t$ memory game, compared to a trivial coin-flip strategy, is bounded, for all $\delta>0$, by
\begin{equation}
H_\infty^{\negl{t}}(X_{[\textit{sample}]}|E)\ge \frac{t\cdot (k-1)}{k} -t\cdot \delta \,.
\end{equation}
\begin{proof}
The result by Vadhan~\cite{vadhan2004constructing} that we used in the classical case has been generalized in presence of quantum side information by K\"onig and Renner in~\cite{konig2011sampling}. Apart from the exact parameter values hidden behind $\negl{t}$, our proof is, hence, unchanged by the addition of quantum side information.
\end{proof}
\end{proposition}

\subsection{The classical SKE protocol is already quantum-resistant}
The information-theoretical security of the SKE protocol from Section~\ref{secske} depends uniquely on the one of privacy amplification and on Theorem~\ref{twothm}.

Since in presence of quantum side information, universal-2 hashing (Lemma~\ref{leftover}) remains a universally composably secure way of achieving privacy amplification~ \cite{renner2005universally,tomamichel2011leftover}, and that, as we just argued, so is the case of Theorem~\ref{twothm}, the SKE scheme presented in Section~\ref{protocolske} is secure against quantum adversaries.

Fundamentally different from standard quantum key distribution, the result is nevertheless an information-theoretically secure key distribution scheme for a quantum world in which entropy is exponentially cheaper than free energy.

\subsection{A quantum-resistance patch for the OT protocol}\label{patchedprotocol}
\begin{figure}[h!]
\begin{center}
\hspace{-1em}\includegraphics[width=0.55\textwidth]{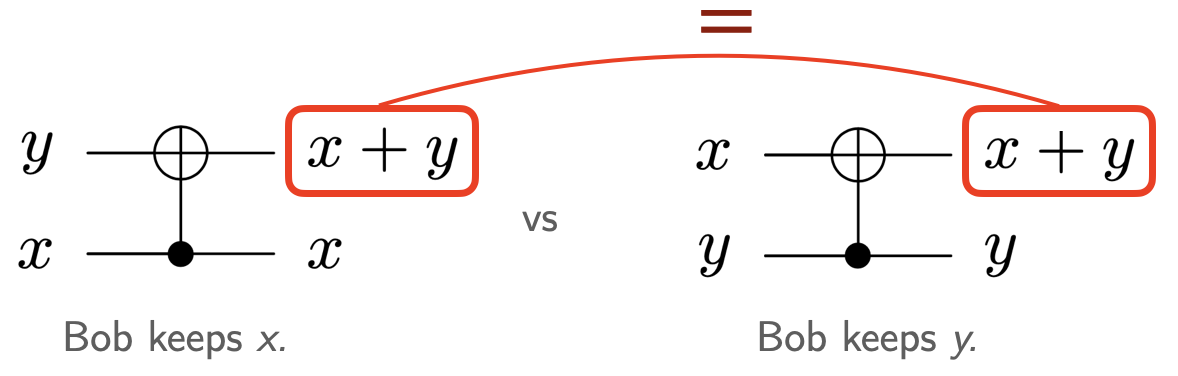}
\caption{If Bob receives a classical state, the top state, $x+y$, that he will return to Alice during the OT protocol will be the same no matter whether he chooses to decrypt the first (left) or second message (right).}
\label{figurexy}
\end{center}
\end{figure}

\begin{figure}[h!]
\begin{center}
\hspace{-2em}\includegraphics[width=0.9\textwidth]{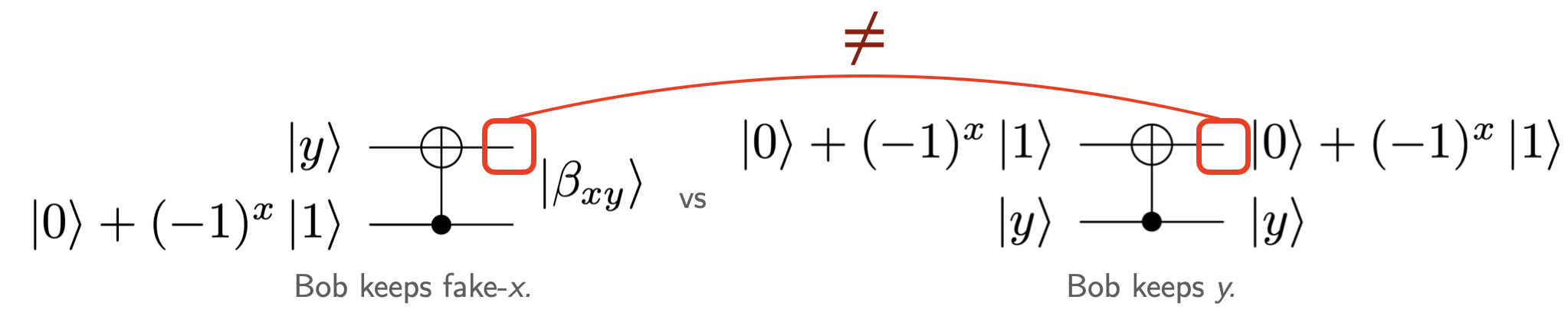}
\caption{A malicious Alice can send to Bob one of the quantum states in the Hadamard basis. In that case, the upper state sent back to Alice by an honest Bob will be $\ket{+}$ or $\ket{-}$ if he wants to keep the first message, but half of one of the four Bell states $\{\ket{\beta_{xy}}\}_{xy}$ if he wants to keep the second message. Since Alice can distinguish between those two cases, the OT scheme is not secure for Bob. Below, we explain how Bob can prevent this quantum attack.}
\label{figurexpasy}
\end{center}
\end{figure}

Given that the above SKE protocol is quantum-resistant, and that the same argument applies to the security-for-Alice part of our oblivious-transfer protocol, it would be natural for our previously detailed scheme to be also quantum-resistant. But it is not: The security-for-Bob, which is trivial in the classical case (because $x+y=y+x$, see Fig.~\ref{figurexy}), can be broken by a malicious quantum Alice. The reason is that if Alice acts maliciously and sends the superposed quantum states $X^{(0)}=H\ket{x}$ and $Y^{(0)}=\ket{y}$ to Bob (for some random $x$ and $y$), she can discriminate between the state sent back by Bob when he does $H\ket{x} \overset{\rm CNOT}{\longrightarrow} \ket{y} $ (to keep $X^{(0)}$) compared to when he does $\ket{y} \overset{\rm CNOT}{\longrightarrow} H\ket{x}$ (to keep $Y^{(0)}$). This attack is illustrated in Fig.~\ref{figurexpasy}.


But there is a simple patch for this attack, or, in fact, for all quantum attacks by a malicious Alice. Alice's extra power comes from the fact she can send states in superposition, but Bob can in return pre\"emptively ``classicize'' the possibly quantum states ${X^{(0)}}$ and ${X^{(1)}}$ by CNOT-ing each bit to a different bit of the totally mixed environments $\pi_0$ and $\pi_1$. Given control of a large enough environment (of dimension $2^{\len{X^{(0)}}+\len{X^{(1)}}}$), Bob can do so at no free energy cost. The resulting state, when traced over that environment, is then undistinguishable from a (possibly noisy) state sent by a malicious-but-classical Alice. Even if misbehaviour from Alice's part might affect the protocol's correctness (which is allowed for a malicious Alice), it leaves the perfect security intact: a quantum Alice can still not gain any information about Bob's choice.

\begin{tcolorbox}[pad at break*=1mm,
  colback=gray!10!white,colframe=gray!75!black,title={Quantum-safe 1--2 oblivious-transfer protocol}]

Steps~1--3~and~5--7 are the same as in the previous classical protocol. Step~4 is changed to

\begin{enumerate} \setcounter{enumi}{3}
\item[4'.] Bob chooses $i\in \{0,1\}$ and computes reversibly
\begin{align*} (X^{(0)},X^{(1)},\pi_0,\pi_1) \rightarrow (X^{(i)},X^{(0\oplus1)},\pi_0 \oplus{X^{(1)}},\pi_1 \oplus{X^{(2)}})\,,
\end{align*} where $\pi_0$ and $\pi_1$ are completely mixed states of appropriate size taken from Bob's environment, and where we define $X^{(0\oplus 1)}\coloneqq X^{(0)}\oplus X^{(1)}$. Then, Bob keeps everything but $X^{(0\oplus1)}$, which he sends back (thermodynamically reversibly) to Alice using \textsc{SWAP}.
\end{enumerate}
\end{tcolorbox}

The above step reduces the security for Bob in the quantum case to the one of the classical case. The updated protocol does not require the honest players to make any quantum operations \emph{per se}.

\section{Concluding remarks}
We propose a \emph{free-energy-bounded} model of cryptography, in which we have derived information-theoretically secure protocols for secret-key establishment and oblivious transfer.

Even if the rationale behind its security is totally different: Our secret-key-establishment protocol is similar to standard quantum key distribution. Our oblivious-transfer protocol, on the other hand, is novel in itself: The mechanism that allows Alice to check that Bob honestly forgets information is proper to reversible computing.

Our schemes are not practical at this point: Current technology is still far from computing with memories that are large enough for Landauer's principle to become the main obstacle (it is worth noting that Boltzmann's constant, which we have in this work conveniently set to $k_{\rm B}:=1/{\rm T}$, is in fact $\approx 1.38\cdot 10^{-23} {\rm J K}^{-1}$); and whereas no laws of physics forbid it, implementing reversible computation on such states is for now science fiction. Our result is rather to be seen as part of the quest of distinguishing what physical phenomena allow for realizing cryptographic functionalities in principle, and which do not. In this spirit, our protocols add another element to the longer and longer list of physical laws from which cryptographic security \emph{can} directly be derived:
We can now claim that information-theoretic key agreement
is theoretically possible as soon as one of the fundamental limits conjectured by \emph{either} quantum theory \emph{or} special relativity \emph{or} \emph{the second law of thermodynamics} is correct.
Concerning the novel appearance of a thermodynamic law in this list, we remark first that according to \emph{Albert Einstein},
thermodynamics is the only physical theory that will survive future development in Physics. Second, the second law is
rather pessimistic in nature, and to see it being linked to a constructive application is refreshing.
We are, in fact, not aware of many uses, besides our protocols, of the law. In summary, we can say, somewhat ironically: \emph{One small step for cryptography --- one giant leap for the second law.}

\subsubsection*{Acknowledgement}

We thank Charles Alexandre B\'edard and Cecilia Boschini for helpful comments about earlier versions of this work.
We also thank Renato Renner and two anonymous referees for some interesting remarks regarding the physicality of the model.
This research was supported by the Swiss National Science Foundation (SNF).

\bibliographystyle{alpha}
\bibliography{../ddd_all.bib}

\end{document}